\documentclass[12pt, twocolumn, numberedappendix, tighten]{aastex631}
\usepackage{savesym}
\savesymbol{tablenum}
\usepackage{natbib}
\usepackage{color} 
\usepackage{morefloats}
\usepackage{gensymb} 
\usepackage{longtable}
\usepackage{amsmath}
\usepackage{graphicx}
\usepackage{subfigure}
\usepackage{siunitx}
\usepackage{appendix}

\newcommand{\kms}{\ifmmode {\rm km\ s}^{-1} \else km s$^{-1}$\fi}
\newcommand{\ergs}{\ifmmode {\rm erg\ s}^{-1} \else erg s$^{-1}$\fi}
\newcommand{\ergscm}{\ifmmode {\rm erg\ s}^{-1} \else erg s$^{-1}$ cm$^{-2}$\fi}
\newcommand{\Msun}{\ifmmode {\rm M}_{\odot} \else $M_{\odot}$\fi }
\newcommand{\Lsun}{\ifmmode {\rm L}_{\odot} \else L$_{\odot}$\fi}
\newcommand{\qo}{\ifmmode q_{\rm o} \else $q_{\rm o}$\fi}
\newcommand{\Ho}{\ifmmode H_{\rm o} \else $H_{\rm o}$\fi}
\newcommand{\ho}{\ifmmode h_{\rm o} \else $h_{\rm o}$\fi}

\newcommand{\vFWHM}{\ifmmode v_{\mbox{\tiny FWHM}} \else
                    $v_{\mbox{\tiny FWHM}}$\fi}
\newcommand{\CCF}{\ifmmode F_{\it CCF} \else $F_{\it CCF}$\fi}
\newcommand{\ACF}{\ifmmode F_{\it ACF} \else $F_{\it ACF}$\fi}
\newcommand{\Halpha}{\ifmmode {\rm H}\alpha \else H$\alpha$\fi}
\newcommand{\Hbeta}{\ifmmode {\rm H}\beta \else H$\beta$\fi}
\newcommand{\Hgamma}{\ifmmode {\rm H}\gamma \else H$\gamma$\fi}
\newcommand{\Hdelta}{\ifmmode {\rm H}\delta \else H$\delta$\fi}
\newcommand{\Lya}{\ifmmode {\rm Ly}\alpha \else Ly$\alpha$\fi}
\newcommand{\Lyb}{\ifmmode {\rm Ly}\beta \else Ly$\beta$\fi}
\newcommand{\HeI}{\ifmmode {\rm He}\,{\sc i}\,\lambda5876 \else 
	          He\,{\sc i}\,$\lambda5876$\fi}
\newcommand{\HeII}{\ifmmode {\rm He}\,{\sc ii}\,\lambda4686 \else 
	           He\,{\sc ii}\,$\lambda4686$\fi}
\newcommand{\heii}{He\,{\sc ii}}
\newcommand{\mgii}{Mg\,{\sc ii}}

\newcommand{\ciii}{\ifmmode {\rm C}\,{\sc iii} \else C\,{\sc iii}\fi}
\newcommand{\civ}{C\,{\sc iv}\ }

\newcommand{\aliii}{Al\,{\sc iii}}
\newcommand{\siiv}{Si\,{\sc iv}}
\newcommand{\nv}{N\,{\sc v}}

\shortauthors{Wheatley et al.}

\begin{document}

\title{The SDSS-V Black Hole Mapper Reverberation Mapping Project: \civ BAL Acceleration in the Quasar SBS~1408+544}

\author{Robert Wheatley}
\affiliation{Department of Astronomy, University of Wisconsin-Madison, Madison, WI 53706, USA} 

\author[0000-0001-9920-6057]{Catherine J. Grier}
\affiliation{Department of Astronomy, University of Wisconsin-Madison, Madison, WI 53706, USA} 
\affiliation{Steward Observatory, The University of Arizona, 933 North Cherry Avenue, Tucson, AZ 85721, USA} 

\author[0000-0002-1763-5825]{Patrick B. Hall}
\affiliation{Department of Physics \& Astronomy, York University, 4700
Keele St., Toronto, ON M3J 1P3, Canada}

\author[0000-0002-0167-2453]{W.~N.~Brandt}
\affiliation{Department of Astronomy and Astrophysics, Eberly College of Science, The Pennsylvania State University, 525 Davey Laboratory, University Park, PA 16802}
\affiliation{Institute for Gravitation \& the Cosmos, The Pennsylvania State University, University Park, PA 16802}
\affiliation{Department of Physics, The Pennsylvania State University, University Park, PA 16802, USA}

\author[0009-0008-3346-3577]{Jonah Lotz} 
\affiliation{Steward Observatory, The University of Arizona, 933 North Cherry Avenue, Tucson, AZ 85721, USA} 

\author{D.~P.~Schneider}
\affiliation{Department of Astronomy and Astrophysics, Eberly College of Science, The Pennsylvania State University, 525 Davey Laboratory, University Park, PA 16802}
\affiliation{Institute for Gravitation \& the Cosmos, The Pennsylvania State University, University Park, PA 16802}

\author[0000-0002-1410-0470]{Jonathan~R.~Trump}
\affiliation{Department of Physics, University of Connecticut, 2152 Hillside Road, Unit 3046, Storrs, CT 06269, USA}

\author[0000-0003-1659-7035]{Yue~Shen}
\affiliation{Department of Astronomy, University of Illinois at Urbana-Champaign, Urbana, IL 61801, USA}
\affiliation{National Center for Supercomputing Applications, University of Illinois at Urbana-Champaign, Urbana, IL 61801, USA}

\author{Lucas~M.~Seaton} 
\affiliation{Department of Physics \& Astronomy, York University, 4700
Keele St., Toronto, ON M3J 1P3, Canada} 

\author[0000-0002-6404-9562]{Scott F. Anderson}
\affiliation{Astronomy Department, University of Washington, Box 351580, Seattle, WA 98195, USA} 

\author[0000-0001-8433-550X]{Matthew J. Temple}
\affiliation{Instituto de Estudios Astrof\'{\i}sicos, Facultad de Ingenier\'{\i}a y Ciencias, Universidad Diego Portales, Avenida Ejercito Libertador 441, Santiago, Chile}

\author{Roberto Assef} 
\affiliation{Instituto de Estudios Astrof\'{\i}sicos, Facultad de Ingenier\'{\i}a y Ciencias, Universidad Diego Portales, Avenida Ejercito Libertador 441, Santiago, Chile}

\author[0000-0001-8032-2971]{Logan B. Fries}
\affiliation{Department of Physics, University of Connecticut, 2152 Hillside Road, Unit 3046, Storrs, CT 06269, USA}

\author[0000-0002-0957-7151]{Y. Homayouni}
\affiliation{Department of Astronomy and Astrophysics, Eberly College of Science, The Pennsylvania State University, 525 Davey Laboratory, University Park, PA 16802}
\affiliation{Institute for Gravitation \& the Cosmos, The Pennsylvania State University, University Park, PA 16802}

\author[0000-0002-2603-2639]{Darshan Kakkad}
\affiliation{Space Telescope Science Institute, 3700 San Martin Drive, Baltimore, MD 21210, USA}

\author[0000-0002-6610-2048]{Anton M. Koekemoer}
\affiliation{Space Telescope Science Institute, 3700 San Martin Drive, Baltimore, MD 21210, USA}

\author[0000-0002-7843-7689]{Mary Loli Mart\'inez-Aldama}
\affiliation{Astronomy Department, Universidad de Concepci\'on, Barrio Universitario S/N, Concepci\'on 4030000, Chile}

\author[0000-0002-1656-827X]{C. Alenka Negrete}
\affiliation{Conacyt Research Fellow at Instituto de Astronom\'ia, Universidad Nacional Aut\'onoma de M\'exico, AP 70-264, CDMX 04510, Mexico}

\author[0000-0001-5231-2645]{Claudio Ricci}
\affiliation{Instituto de Estudios Astrof\'{\i}sicos, Facultad de Ingenier\'{\i}a y Ciencias, Universidad Diego Portales, Avenida Ejercito Libertador 441, Santiago, Chile}
\affiliation{Kavli Institute for Astronomy and Astrophysics, Peking University, Beijing 100871, China}

\author[0000-0002-3601-133X]{Dmitry Bizyaev}
\affiliation{Apache Point Observatory and New Mexico State
University, P.O. Box 59, Sunspot, NM, 88349-0059, USA}
\affiliation{Sternberg Astronomical Institute, Moscow State
University, Moscow}

\author{Joel~R.~Brownstein} 
\affiliation{Department of Physics and Astronomy, University of Utah, 115 S. 1400 E., Salt Lake City, UT 84112, USA}

\author[0000-0002-6770-2627]{Sean~Morrison} 
\affiliation{Department of Astronomy, University of Illinois at Urbana-Champaign, Urbana, IL 61801, USA}

\author{Kaike Pan}
\affiliation{Apache Point Observatory and New Mexico State
University, P.O. Box 59, Sunspot, NM, 88349-0059, USA}

%%%%%%%%%%%%%%%%%%%%%%%%%%%%%%%%%%%%%%%%%%%BEGIN PAPER %%%%%%%%%%%%%%%%%%%%%%%%%%%%%%%%%%%%%%%%%%%%%%%%%%%%%%%%%%%%%%%%%%%

\begin{abstract}
We present the results of an investigation of a highly variable \civ  broad absorption-line feature in the quasar SBS 1408+544 ($z = 2.337$) that shows a significant shift in velocity over time. This source was observed as a part of the Sloan Digital Sky Survey Reverberation Mapping Project and the SDSS-V Black Hole Mapper Reverberation Mapping Project, and has been included in two previous studies, both of which identified significant variability in a high-velocity \civ broad absorption line (BAL) on timescales of just a few days in the quasar rest frame. Using $\sim$130 spectra acquired over eight years of spectroscopic monitoring with SDSS, we have determined that this BAL is not only varying in strength, but is also systematically shifting to higher velocities. {Using cross-correlation methods, we measure the velocity shifts (and corresponding acceleration) of the BAL on a wide range of timescales, measuring an overall velocity shift of $\Delta v$~=~$-683^{+89}_{-84}$~km~s$^{-1}$ over the 8-year monitoring period}. This corresponds to an average rest-frame acceleration of $a$~=~1.04$^{+0.14}_{-0.13}$~cm~s$^{-2}$, though the magnitude of the acceleration on shorter timescales is not constant throughout. We place our measurements in the context of BAL-acceleration models and examine various possible causes of the observed velocity shift. 

\end{abstract}


\section{INTRODUCTION}
\label{introduction}

Broad absorption lines (BALs) in the spectra of quasars are thought to originate in winds that are launched from quasar accretion disks (e.g., \citealt{Murray95, Proga00, Higginbottom14, Naddaf23}). BALs are defined as absorption features with velocity widths wider than 2000 \kms\ (\citealt{Weymann91, Hall13}) and are found in roughly 10-15\% of all optically selected quasars (\citealt{Gibson09b, Allen11}). The winds/outflows that produce BALs may play an important role in the evolution of galaxies --- if these outflows are sufficiently energetic, the gas can produce significant feedback that interferes with star formation within the host galaxy and/or further growth of the supermassive black hole (e.g., \citealt{Dimatteo05, Moll07, King10}). Characterizing these BALs and their environments to constrain models for how they are produced, how they evolve, and how they affect their galaxies, is thus important for our understanding of galaxy evolution. 

BALs are seen across a wide variety of ionization species (\citealt{Turnshek84, Arav01}), one of the most common of which is C\,{\sc iv}, which appears as a doublet at rest-frame wavelengths of 1548.20\,\AA\ and 1550.77\,\AA. \civ is one of the higher-ionization BAL species that appears the most often in quasar spectra, and is high enough in abundance that it is often saturated. It is thus often probed using \siiv\ as a tracer, as \siiv\ is much lower abundance and thus usually unsaturated and can be used for density estimations (e.g., \citealt{Arav18}). \cite{Xu19} use photoionization modeling to determine the physical conditions of BAL outflows using \siiv\ and find that high-ionization BALs such as \civ can probe regions of gas with electron densities ranging from 10$^3$ -- 10$^{5.5}$~cm$^{-3}$, column densities of log $N_{\rm H}$ = 20 -- 22.5, and log $U_{\rm H}$ between $-2$ and 0 (see their Figure 4); in addition, their photoionization models return a range of temperatures for these outflows ranging from 15,000 K to 20,000 K. Many of the studies that have used photoionization modeling to determine the distance of these outflows find that the outflows are at distances of $>$ 500 pc from the central source (see \citealt{Arav18} and references therein).

BALs are variable on rest-frame timescales ranging from days to years (e.g., \citealt{Lundgren07, Gibson08b, Capellupo12, Filizak13, Vivek14, Grier15, Grier16}). The most commonly observed modes of variability are changes in strength and/or profile shape. This observed variability allows us to place constraints on the geometry, distance, and dynamics of the outflows themselves, possibly revealing information on the energetics of the outflows to inform models of quasar/host-galaxy feedback (e.g., \citealt{Arav13}).
In addition to changes in equivalent width and profile shape, monolithic velocity shifts, or ``acceleration", of BAL outflows have also been reported. The presence (or lack) of observed acceleration in BALs provides an important test for models describing the production and evolution of BAL outflows, as some of these models predict visible acceleration (e.g., \citealt{Higginbottom14}). 

Over the past two decades, there have been a handful of reports of observed velocity shifts (implying BAL acceleration or deceleration) in several different studies (\citealt{Vilkoviskij01, Rupke02, Gabel03, Hall07, Joshi14, Grier16, Joshi19, Lu19, Lu20, Xu20, Byun22}). However, velocity shifts caused by an increase or decrease in speed of outflowing gas are difficult to identify. First, because BALs are quite variable in line profile, it can be difficult to differentiate between shifts caused by velocity-dependent line-profile variability (due to changes in the ionization state of the absorbing gas or changes in the column density coverage, for example) and shifts caused by an actual change in the speed of the outflow. In addition, previously measured velocity shifts have mostly been small in magnitude over short timescales, with typical velocity shifts of only a few hundred km~s$^{-1}$ over a few years in the quasar rest frame. We thus expect to require long time baselines to observe a significant velocity shift. 

The Sloan Digital Sky Survey Reverberation Mapping Project (SDSS-RM; \citealt{Shen15a}) and the SDSS-V Black Hole Mapper Reverberation Mapping Program (BHM-RM; \citealt{Kollmeier19}) provide us with an excellent opportunity to explore BAL variability, and acceleration, in quasars. The SDSS-RM program observed a single field of $\sim$850 quasars from 2014--2020; these observations began as a part of the SDSS-III and SDSS-IV surveys (\citealt{Eisenstein11, Blanton17}). Roughly 380 of these quasars continue to be monitored as a part of the SDSS-V BHM-RM program, further extending the time baseline of these observations. While the primary goal of the SDSS-RM and BHM-RM monitoring programs is to measure black-hole masses using the technique of reverberation mapping, there are roughly 90 quasars in this sample that show BAL features. With more than 100 spectral observations of these quasars over 8 years (and counting), this survey is very well suited for studies of BAL variability, allowing us to explore it on both short (few-day; e.g., \citealt{Hemler19}) and long (several-year) timescales. 

One of the sources observed by the SDSS-RM and BHM-RM programs is the quasar SBS 1408+544 (SDSS\,J141007.72+541203.6, hereafter referred to as RM\,613). This quasar has a redshift of $z~=~2.337~\pm~0.003$ (\citealt{Shen19}), an apparent $i$-band magnitude $m_i~=~18.1$ (\citealt{Alam15}), and an absolute magnitude $M_i~=~-27.69$. This source was somewhat serendipitously discovered to show strong variability in the equivalent width of its \civ BAL on very short timescales --- \cite{Grier15} studied the first 32 observations of this quasar during the first year of monitoring (each observation was taken on a different night, hereafter referred to as an epoch) and found that the equivalent width of the BAL was changing significantly on timescales down to just 1.20 days in the quasar rest-frame. \cite{Hemler19} also included RM 613 in their sample study of \civ\ BAL variability in $\sim$30 SDSS-RM BAL quasars, which included four years of SDSS-RM monitoring. They confirmed the short-term variability that was originally reported by \cite{Grier15}, but in the subsequent three years of observations, this BAL weakened somewhat and did not show additional dramatic variability. We have since acquired roughly 70 additional spectra of the SDSS-RM field over four additional years of monitoring by SDSS, so we are following up on all of the sources that were found to have significant short-timescale variability by \cite{Hemler19} to investigate whether they continued to show strong variability. We here revisit the quasar RM\,613 as a part of this followup effort. 

In Section~\ref{sec:data}, we discuss the observations, the preparation of the spectra for analysis, and our continuum-normalization procedure. Section~\ref{sec:measurements} describes the measurements made and the tests we performed, and Section~\ref{sec:discussion} includes a discussion of these results, their implications, and their relevance to models of BALs. Where necessary, we adopt a flat cosmology with $H_0$=~70~km~s${-1}$~Mpc$^{-1}$, $\Omega_M$~=~0.3 and $\Omega_{\Lambda}$~=~0.7. 

\section{DATA AND DATA PREPARATION} 
\label{sec:data}

\subsection{Spectral Data}
\label{sec:spectra}
\begin{figure*}
\begin{center}
\includegraphics[scale = 0.3, trim = 0 0 0 0, clip]{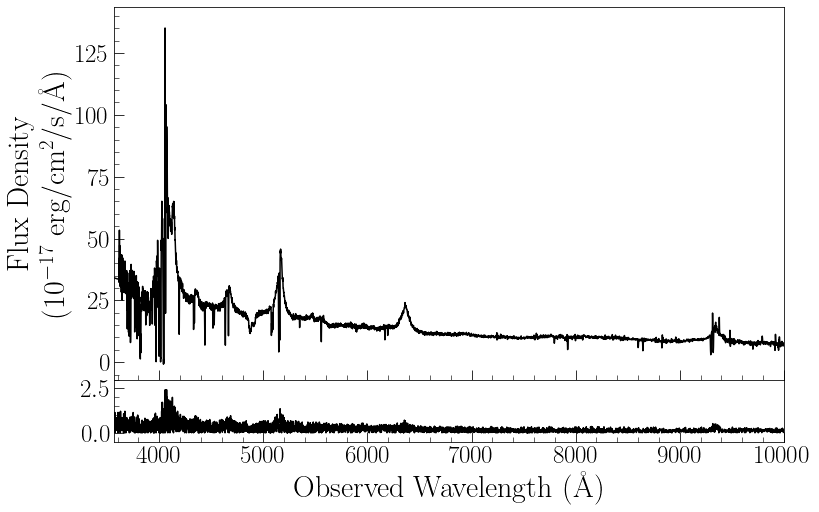}
\includegraphics[scale = 0.3, trim = 0 0 0 0, clip]{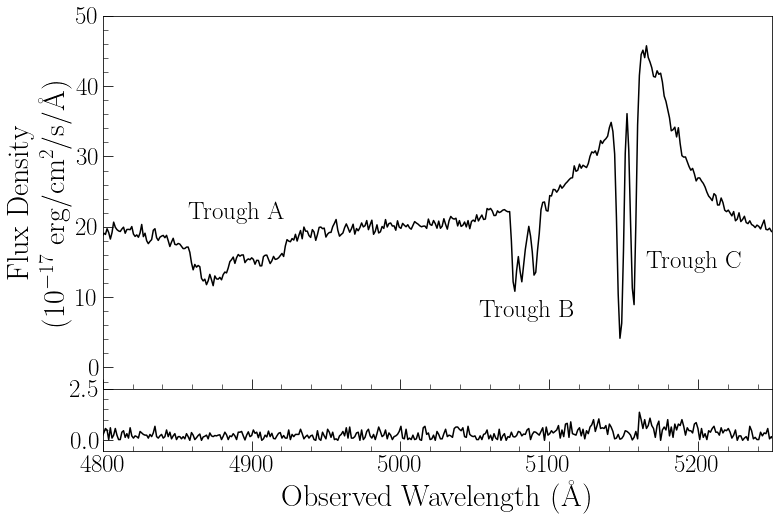}
%mean_raw_spec.png, meanbal_raw_spec.png
\caption{The mean spectrum of RM\,613, calculated from the 129 SDSS spectral epochs used in our study (see Section~\ref{sec:dataprep}). The left panel shows the entire wavelength range covered by the SDSS spectra; the right panel is zoomed in on the \civ region of interest. {The top panels show the flux density in units of 10$^{-17}$~erg~s$^{-1}$~cm$^{-2}~$\AA$^{-1}$ and the bottom panels show the spectral uncertainties in the same units. }}
\label{fig:mean_raw_spectrum} 
\end{center} 
\end{figure*}

The spectra used in this study are from the SDSS-RM project (e.g., \citealt{Shen15a}), which was carried out as a part of the SDSS-III and SDSS-IV programs (\citealt{Eisenstein11, Blanton17}) from 2014--2020. In addition, we obtained 41 spectra of this source as a part of the SDSS-V BHM-RM program (\citealt{Kollmeier19}), which began observations in 2021. There is also an additional spectrum of this source taken as a part of the SDSS-III survey in 2013~May (before the SDSS-RM program began) that was released to the public as a part of the SDSS Data Release 12 (DR12; \citealt{Alam15}); we include this spectrum in our study as well. In total, we have 132 spectra of this quasar taken with the 2.5-m SDSS telescope at Apache Point Observatory with the BOSS spectrograph (\citealt{Gunn06, Dawson13, Smee13}) spanning roughly nine years (eight years of dedicated monitoring, plus the early SDSS spectrum taken a year earlier). The BOSS spectrograph covers a wavelength range of roughly 3650--10,400\,\AA\ with a spectral resolution of $R~\sim~2000$, which results in a velocity resolution of $\sim$69 km~s$^{-1}$ in the \civ region of the spectrum. The spectra from the first two years of monitoring (2013 and 2014) were processed with the standard SDSS-III pipeline (version $5\_7\_1$) and the remaining years of data were processed with the updated SDSS-IV eBOSS pipeline (version $5\_10\_1$). Figure~\ref{fig:mean_raw_spectrum} shows the mean spectrum of RM\,613, created from these spectra. 

\begin{figure}
\begin{center}
\includegraphics[scale = 0.3, trim = 0 0 0 0, clip]{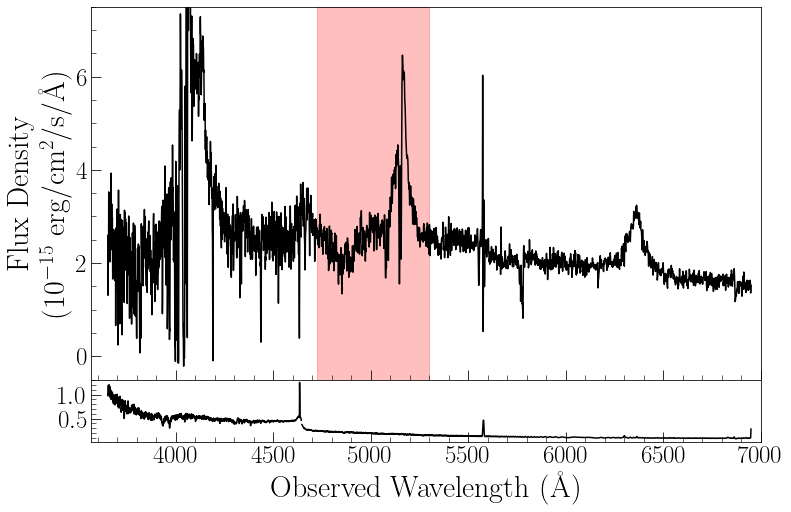} %het_raw_spec.png 
\caption{The spectrum of RM\,613 taken on 30 May 2023 by the Hobby-Eberly Telescope (HET). {The top panels show the flux density in units of 10$^{-15}$~erg~s$^{-1}$~cm$^{-2}$~\AA$^{-1}$ and the bottom panels show the spectral uncertainties in the same units. The red shaded region indicates the \civ region of interest. The HET spectrum has been rebinned to the same wavelengths as the SDSS spectra using the {\tt SpectRes} python package for display purposes.}}
\label{fig:het_raw} 
\end{center} 
\end{figure}

With the recent upgrade to the SDSS-V robotic fiber positioning system (FPS), our original field of 849 quasars had to be reduced to approximately 380 sources for continuing observations. RM\,613 was dropped from the SDSS-V monitoring because its position in the SDSS-RM field is outside the FPS field of view, so we have not been able to continue monitoring this particular source with SDSS-V. However, we were able to obtain a spectrum of RM\,613 on 30 May 2023 with the Hobby-Eberly Telescope (\citealt{Ramsey98, Hill21}) using the Low Resolution Spectrograph 2 (LRS2; \citealt{Chonis16}). Figure~\ref{fig:het_raw} shows that the \civ BAL in question has weakened significantly at this point. This spectrum was processed separately from the SDSS spectra; see Appendix~\ref{app:het} for details.   

We also note that an additional spectrum of this source was taken in 1991 by \cite{Chavushyan95}. \cite{Grier15} inspected this spectrum and noted that  the high-velocity \civ BAL was not detectable in the spectrum at the time, though the SNR was too low for its absence to be established definitively. We do not include this spectrum in our analysis, as the signal is too low for useful measurements to be obtained. 

Visual inspection of the mean spectrum (Figure~\ref{fig:mean_raw_spectrum})  confirms that this quasar hosts a prominent high-velocity \civ BAL feature that is detached from the \civ emission line. Following \cite{Grier15}, we will henceforth refer to this BAL as ``Trough A". In addition to this higher-velocity \civ BAL, we see two narrower \civ absorption systems that are superimposed onto the \civ emission line (hereafter ``Trough B" for the middle feature and ``Trough C" for the narrower feature that lies nearly at line center; see Figure~\ref{fig:mean_raw_spectrum}). We see no BALs in lower-ionization transitions (e.g., \mgii\ or \aliii), which means this source is considered a ``high-ionization" BAL quasar. There is a hint of \siiv\ absorption at similar velocities as \civ but the absorption is too weak to formally meet the definition of a BAL. While it is visible in the mean spectrum, this absorption is too shallow for us to well measure the properties of this feature in individual spectra. In addition to the difficulties presented by the shallowness of the line, its blue edge is contaminated by O{\sc i}/ Si {\sc ii} emission, and there is a strong, narrow absorption line located on the red edge. Both of these features interfere with attempts to make measurements of the \siiv\ absorption even in mean/stacked spectra. We also see some \nv\ absorption just redward of \Lya, but it is too contaminated by \Lya\ absorption for us to include in our study.

\subsection{Data Preparation}
\label{sec:dataprep}  

Prior to any further processing, we visually inspected all spectra and noticed a few epochs with significant excess noise. We quantify the SNR of each spectrum at rest-frame 1700\,\AA\ by measuring the median SNR of pixels between rest-frame 1650--1750\,\AA\ (SNR$_{1700}$), which corresponds to 5506--5840\,\AA\ in the observed frame for RM\,613. Based on visual inspection, we determined that spectra with SNR$_{1700}~<~3$ were unlikely to provide any useful constraints, so we excluded all epochs with SNR$_{1700}$ below this threshold. This excluded three epochs: 7, 53, and 115. This brings the total number of usable BOSS spectra down to 129 (including the early BOSS spectrum). 

We first cropped off all pixels at wavelengths less than 3650\,\AA\ and greater than 10300\,\AA\ (in the observed frame), as both the blue and red edges of the spectra showed significant noise and telluric contamination. We then searched for pixels that may have problems, as flagged by the SDSS pipeline using the bitmasks provided with the spectra. We linearly interpolated over any pixels that were flagged by the SDSS ``ANDMASK" as having issues. The uncertainties on the interpolated pixels were multiplied by a factor of 10 to represent the increased uncertainties due to interpolation. 
We follow previous work (e.g. \citealt{Gibson09b, Grier15}) and correct for Galactic extinction and reddening in the spectra, adopting a $R_V = 3.1$ Milky Way extinction model (\citealt{Cardelli89}) and $R_V$ values following \cite{Schlafly11}. We then shifted the spectra to the quasar rest frame using a redshift of $z = 2.337$ (\citealt{Grier15}). All further discussion/analysis of these spectra will refer to the spectra in the quasar rest frame. 

\subsection{Continuum and Emission-Line Fits} 
\label{sec:confits} 
\begin{figure*}
\begin{center}
\includegraphics[scale = 0.31, trim = 0 0 0 0, clip]{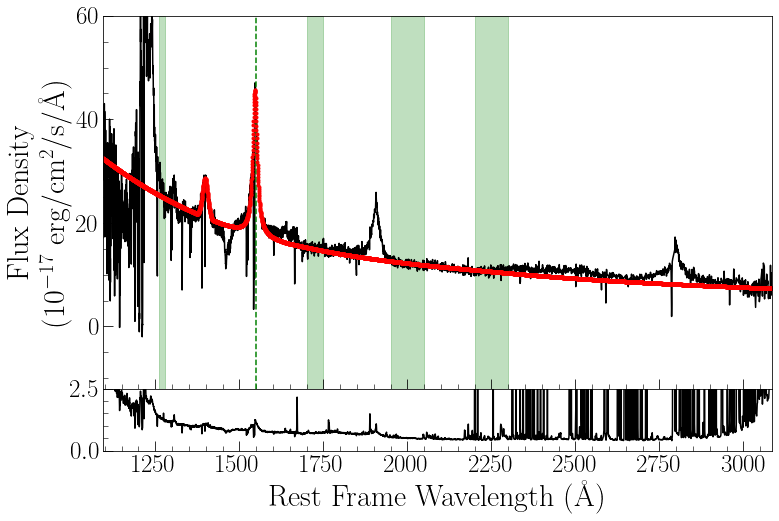} %civ_cont_fit.png
\includegraphics[scale = 0.31, trim = 0 0 0 0, clip]{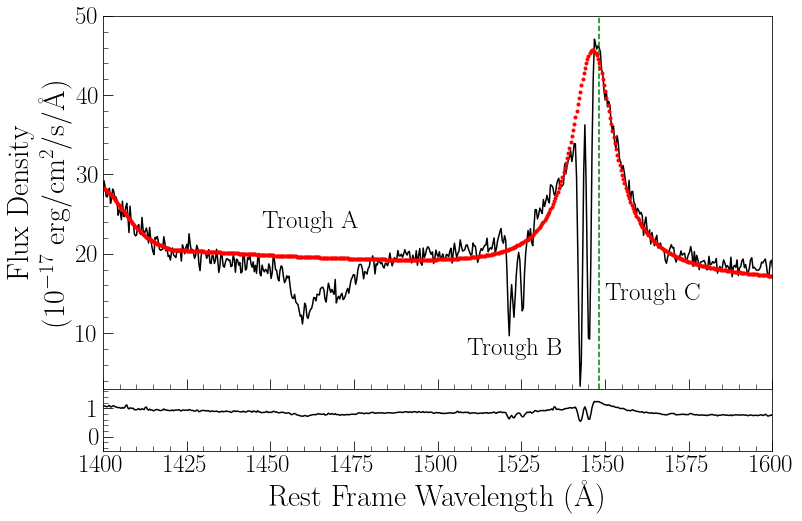} %civ_cont_fit_bal.png
\caption{An example SDSS spectrum of RM\,613 and its corresponding continuum fit. The spectrum itself is shown in black, the continuum fit --- {including the \civ and \siiv\ emission-line fits} --- in red, and the green shaded regions highlight the four line-free regions used in the continuum fit. The vertical dashed green line shows the rest-frame wavelength of C\,{\sc iv}. The right panel shows the same spectrum zoomed in to the \civ region of interest. {The top panels show the flux density in units of 10$^{-17}$~erg~s$^{-1}$~cm$^{-2}$~\AA$^{-1}$ and the bottom panels give the spectral uncertainties in the same units. }} 
\label{fig:con_fit} 
\end{center} 
\end{figure*}

To isolate the variability of the BAL from the variability of the rest of the quasar, we fit a continuum model to each spectrum. We follow previous work (e.g., \citealt{Filizak12}) and model the quasar continuum as a reddened power law with an SMC-like reddening coefficient (\citealt{Pei92}). We fit the continuum using a nonlinear least-squares algorithm, selecting four ``line-free regions" ---  wavelength ranges that are largely uncontaminated by strong emission and absorption features (1270\,\AA\ -- 1290\,\AA, 1700\,\AA\ -- 1750\,\AA, 1950\,\AA\ -- 2050\,\AA, and 2200\,\AA\ -- 2300\,\AA). An example spectrum with its continuum fit is shown in Figure~\ref{fig:con_fit}. We calculated the uncertainties in the continuum fits using ``flux randomization" Monte Carlo iterations, where we altered the flux of each individual pixel by a random Gaussian deviate based on the size of its uncertainty. We then fit the continuum to the new altered spectrum, and repeated this process 100 times to determine the standard deviation of the model continuum flux at each pixel, which we adopt as the uncertainty in the continuum fit. 

The primary BAL of interest in our work, Trough A, is at a sufficiently high velocity that it appears detached from the \civ emission line. However, there are two additional absorption features that lie on top of the \civ emission line; in order to examine their behavior, we need to isolate the \civ emission line from the BALs. To do this, we follow \cite{Hemler19} and fit an emission-line profile to each individual spectrum and divide it out with the continuum. We chose to use a Voigt profile (e.g., \citealt{Gibson09b}), which was found by \cite{Hemler19} to be a good fit in this particular source. We used an iterative fitting technique to exclude wavelength bins where the flux deviated from the fit by more than 3$\sigma$ and manually excluded the regions that showed significant absorption. As expected, the \civ emission line model was not so wide that it affects the high-velocity Trough~A BAL; as such, our primary analysis, which focuses on Trough~A, does not rely at all on our emission-line fits. 

{We also investigated the possibility that the \siiv\,$\lambda$1393 emission line, with line center at 1393.755\,\AA, may interfere with our measurements. We fit a Voigt profile to the \siiv\ line using the same procedure as with \civ, but found that the red wing of the \siiv\ line ended significantly blueward from the blue wing of Trough A; there is no overlap between the \siiv\ emission line and the BAL feature at any epoch. As with the \civ line, the \siiv\ line thus does not affect any of our measurements of Trough A. The \siiv\ fits are shown in Figure~\ref{fig:con_fit} for demonstrative purposes, but are not included in the majority of our analysis. }

\begin{figure}
\begin{center}
\includegraphics[scale = 0.33, trim = 0 0 0 0, clip]{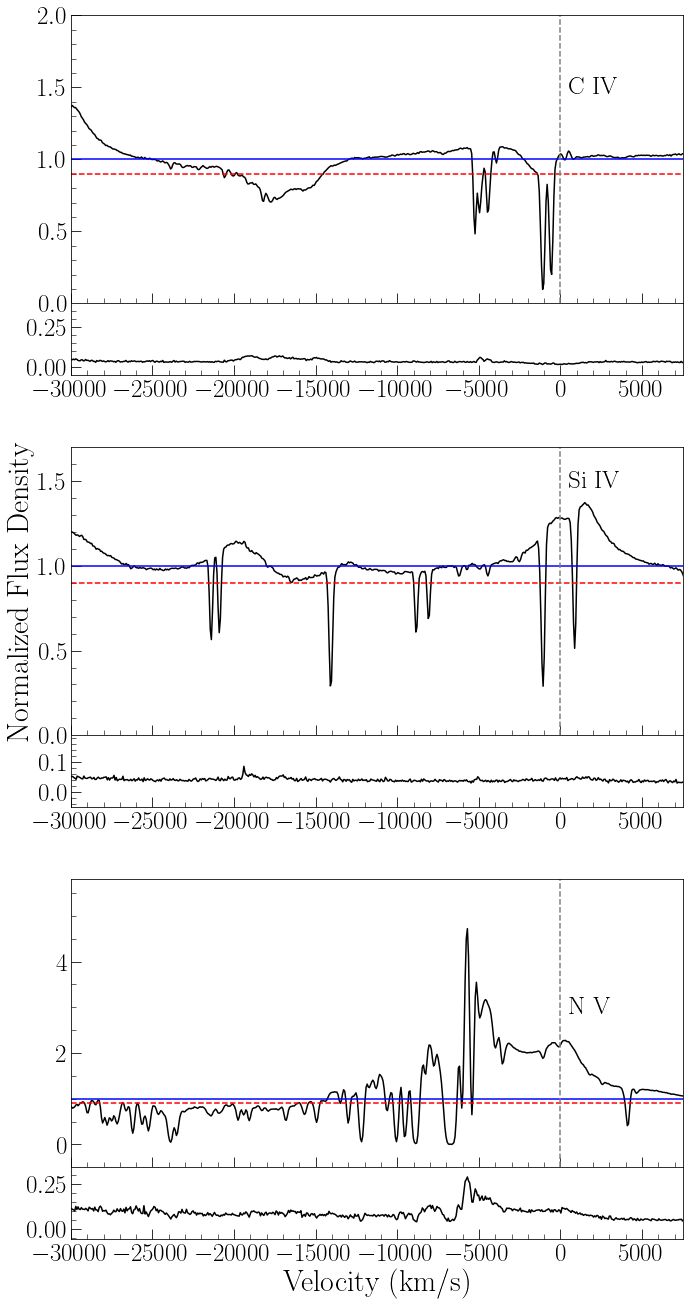}%meanbal_norm_spec.png}
\caption{{Mean spectra used in our study, after normalizing by the continuum and \civ emission-line fit. The three panels are zoomed into three regions of interest (the \civ, \siiv, and \nv\ regions). Dotted vertical lines indicate the rest-frame line center of the three emission lines; the velocity given on the x-axes represents the velocity calculated against each particular emission line. The solid blue horizontal line indicates a normalized flux density of 1.0, and the red dotted horizontal line shows a flux density of 0.9, to aid the eye. The top subpanel for each region shows the mean normalized flux density, and the smaller subpanel shows the uncertainties on the mean normalized spectrum. }}
\label{fig:mean_normalized} 
\end{center} 
\end{figure}

We added the continuum and \civ emission-line models together and divided the original spectra by this combined continuum$+$emission-line fit to obtain a set of ``normalized" spectra. The uncertainties from the continuum fit were propagated along with the spectral uncertainties to determine the final uncertainties on the normalized spectra. All subsequent measurements and analysis were performed on the normalized spectra unless otherwise noted. Figure~\ref{fig:mean_normalized} shows the mean normalized spectrum, focused on the \civ region of the spectrum. Because we only fit the continuum and \civ emission line, other emission features are still visible at the edges of this range (we see \siiv\ around 1400 \AA\ and low-level contributions from a wide variety of species redward of C\,{\sc iv}); however, none of these features has any effect on our measurements of Trough A, as the local continuum around Trough A is well fit by the continuum. In addition to Troughs A, B, and C, there are a number of narrow absorption features present in the spectrum as well, some of which overlap on top of Trough A and B. These have been identified as intervening \civ and \siiv\ systems at a variety of redshifts. 

\section{Variability and Acceleration Measurements} 
\label{sec:measurements}
\subsection{BAL Measurements} 

\begin{figure}
\begin{center}
\includegraphics[scale = 0.28, trim = 0 0 0 0, clip]{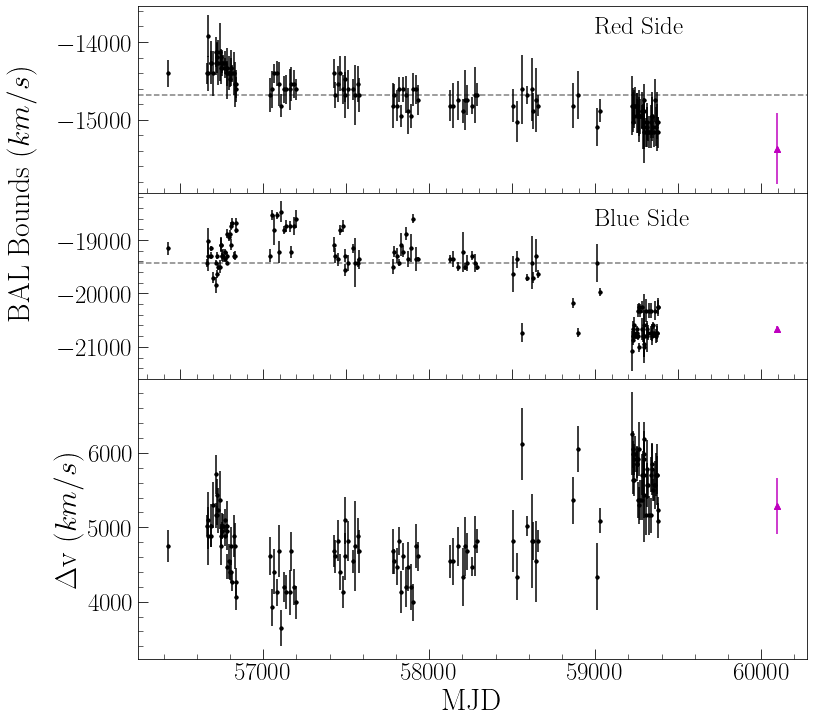} %bounds_dv_mjd.png}
\caption{The redward (top panel) and blueward (middle panel) boundaries of the Trough A BAL as a function of the Modified Julian Date (MJD). Black points represent SDSS spectra, and the magenta triangles represent the measurements made from the HET spectrum. These boundaries are defined as the location at which the flux of the BAL reaches a normalized flux of 0.9 on either side of the BAL feature. Horizontal dashed lines indicate the median value throughout the monitoring, to aid the eye. The bottom panel shows the velocity width ($v_{\rm max} - v_{\rm min}$) of the BAL as a function of time using the determined limits.} 
\label{fig:bal_bounds} 
\end{center} 
\end{figure}

\begin{figure}
\begin{center}
\includegraphics[scale = 0.29, trim = 0 0 0 0, clip]{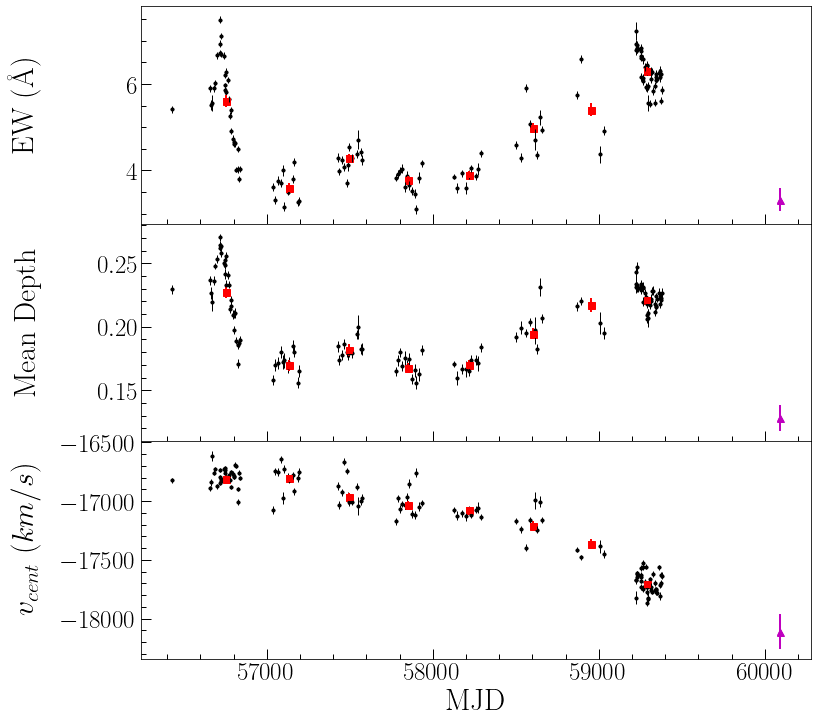} %things_vs_mjd.png}
\caption{Measurements of Trough A over time. The top panel shows the equivalent width (in units of \AA), the second panel shows the mean depth of the BAL, and the third panel shows the absorbed-flux-weighted centroid velocity ($v_{\rm cent}$) of the BAL. Black points represent measurements made from each individual SDSS spectrum and the magenta triangles represent the measurements made from the HET spectrum. Red squares show the values measured from the mean spectra of each individual year (see Section~\ref{sec:meanspectra}).}
\label{fig:things_vs_mjd} 
\end{center} 
\end{figure}

Our first goal was to examine the variability of the \civ BAL Trough A with the additional four years of monitoring with SDSS that were not included by \cite{Hemler19}. To do this, we first characterize the properties of Trough A by measuring its rest-frame {equivalent width (EW, in units of \AA), as defined by
\begin{equation}
	\rm{EW} = \int_{\lambda_{\rm min}}^{\lambda_{\rm max}} [1 - f_{n}(\lambda)]\ \rm{d}\lambda
\end{equation}
where $f_n$ represents the continuum-normalized flux. We also measure the absorbed-flux-weighted velocity centroid in units of km~s$^{-1}$, 
\begin{equation}
	v_{\rm {cent}} = \frac{\int_{v_{\rm min}}^{v_{\rm max}} v [1 - f_{\rm n}(v)]\ \rm{d}\textit{v}}{\int_{v_{\rm min}}^{v_{\rm max}}{[1-f_{\rm n}(v)]}\ \rm{d}\textit{v}}
\end{equation}
and the mean fractional depth ($d$, in units of normalized flux)}. We first smoothed the spectra using a boxcar smoothing algorithm over five pixels; this smoothed spectrum was used to determine the upper and lower wavelength limits of the BALs, defined as the location where the flux in the BAL reaches 90\% of the continuum flux (corresponding to a normalized flux density of 0.9 as per standard BAL definitions/conventions; see, e.g., \citealt{Gibson09b}) on either side of the BAL. Figure~\ref{fig:bal_bounds} shows these boundaries over the course of the monitoring period. We then measured the properties of the BAL within these outer limits, using Monte Carlo randomizations to measure the uncertainties in these parameters. These measurements are shown in Figure~\ref{fig:things_vs_mjd}.  

Figure~\ref{fig:things_vs_mjd} indicates how the BAL varied over the monitoring period using these three measurements. We again confirm the strong short-timescale variability observed by \cite{Grier15} during the first year of observations. This was followed by a period of about 4 years where the BAL remained in a weaker state before it again strengthened. The mean depth shows the same variability as the EW. However, inspection of $v_{\rm cent}$ over time shows a very strong trend toward higher (more negative) velocities over the monitoring period. This could be indicative of acceleration of the BAL. Examination of the upper and lower limits of the BAL as a function of time (Figure~\ref{fig:bal_bounds}) confirms that the upper and lower bounds of the BAL are also slowly shifting blueward over time.

\begin{figure*}
\begin{center}
\includegraphics[scale = 0.4, trim = 0 0 0 0, clip]{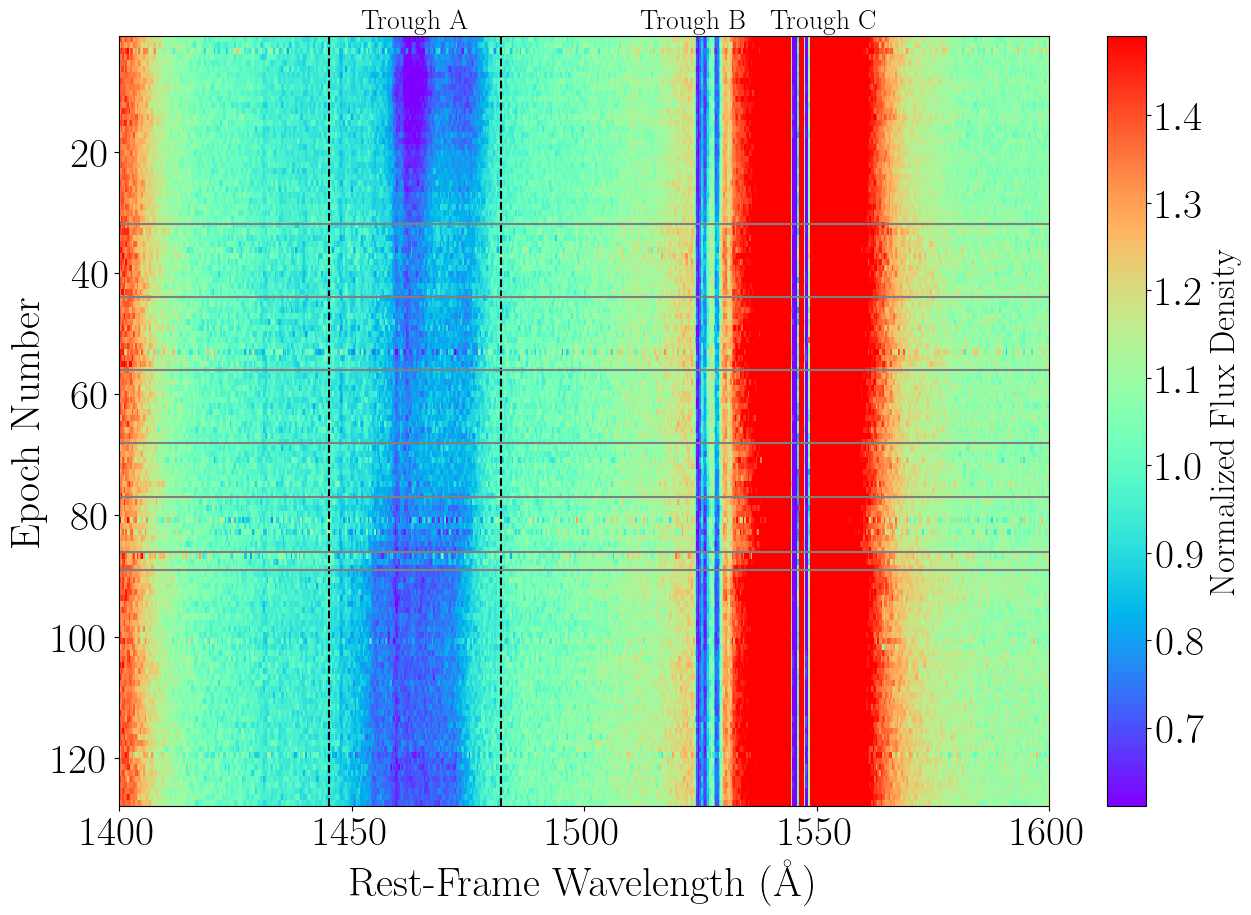}%trailed_spectrogram.png}
\caption{A trailed spectrogram showing all of our continuum-normalized spectra over time. For visualization purposes, we did not include the emission-line fit in the normalization for this figure. Each row represents a different spectrum, and the color indicates the strength of the normalized flux density in each wavelength bin: absorption features appear as cool/blue vertical stripes, and emission lines appear as warm/red stripes. Vertical black dashed lines indicate the highest and lowest boundaries of the \civ BAL throughout the monitoring period. Horizontal gray lines indicate the breaks between observing seasons. Occasional narrow horizontal ``stripes" of noise that appear are spectra that are noticeably lower in SNR than most of the other spectra. It is apparent that Trough A shifts to the left (to higher blueshifted velocities) throughout the monitoring.} 
\label{fig:trailed_spec} 
\end{center} 
\end{figure*}

Figure~\ref{fig:trailed_spec} shows a ``trailed spectrogram" that includes all 129 of the SDSS spectra included in our study as a way of visualizing the normalized flux over time. For visualization purposes, we used the continuum-normalized spectra {\it without} the \civ emission-line fit to produce this figure. Examination of this plot confirms our conclusions based on the measurements of the BAL. First, the blueward shift of the BAL over the course of the observations (as suggested by our $v_{\rm cent}$ vs. time plot) is readily apparent. The trailed spectrogram also indicates the relative weakening of the absorption during the middle 3-4 years of monitoring, followed by a regaining of strength in the final three years of monitoring. A similar trend in strength is observed in the Trough~B absorption feature (appearing at roughly 1535\,\AA), though Trough~C, the narrow doublet feature that lies close to line center, does not appear to vary significantly in strength. {Faint, thin, vertical blue lines that are visible in Figure~\ref{fig:trailed_spec} (particularly those just blueward of Trough A) represent the narrow intervening absorption systems that we identified (Section~\ref{sec:confits})}. 

\begin{figure}
\begin{center}
\includegraphics[scale = 0.25, trim = 30 0 0 0, clip]{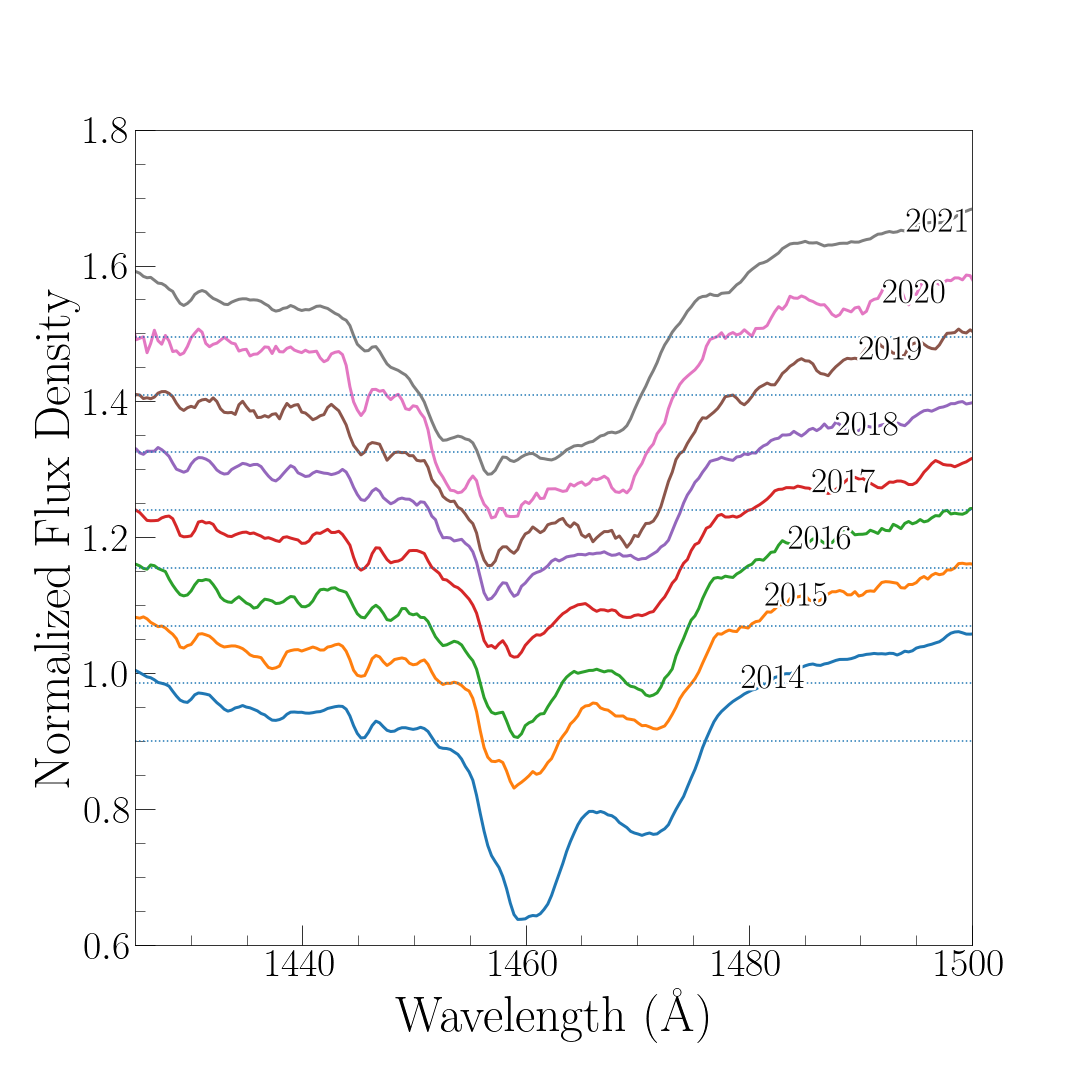}%smooth_balmean_spectra.png}
\caption{Mean spectra from each observing year, focused in on the \civ BAL. The bottom (blue) line shows the mean from 2014, and each line moving upward represents the next year in sequence (2015-2021), with the top curve showing the mean in 2021. These mean spectra have been smoothed by five pixels using a boxcar smoothing algorithm, and each spectrum has been shifted upward in flux for visibility purposes. The horizontal blue dotted lines indicate a normalized flux level of 0.9 for each spectrum to aid the eye in identifying the upper and lower limits of the BAL.} 
\label{fig:mean_specs_by_year} 
\end{center} 
\end{figure}

Despite the significant changes in EW, \cite{Grier15} found no significant line-profile variability on short timescales during the first year of monitoring. However, our Figure~\ref{fig:trailed_spec} indicates a change in the shape of the BAL over the longer time period. To examine how the shape of the BAL changed over the longer time baseline spanned by our data, we combined each year of monitoring, producing a mean spectrum for each individual year, and inspected the shape of the line in these higher-SNR spectra. Figure~\ref{fig:mean_specs_by_year} shows that the average line profile remained similar through the first several years of monitoring; however, over the last 2-3 years of monitoring, the trough changed from having two distinct sub-troughs to a more uniform, single-trough feature. The slow blueward shift in velocity of this BAL is still quite apparent in Figure~\ref{fig:mean_specs_by_year}. Interestingly, our recent HET spectrum (see Appendix~\ref{app:het}) shows that the trough had returned to its two-pronged structure by mid-2023, with Trough A presenting as two distinct troughs where the normalized flux returns to a value of 1 between them (formally, this feature would be identified as two mini-BALs in this spectrum rather than a BAL). 

Previous studies of BAL acceleration typically only include 2-3 spectra, though some studies have had as many as five (\citealt{Byun22}). The more spectra, the more clear it is whether or not the observed shift in velocity is ``real" or not -- i.e., whether the shift may be due to velocity-dependent variability across the trough, or is actually a shift in velocity that could be caused by acceleration of the outflow itself. With 129 epochs spanning 9 years, it is clear that this BAL is undergoing an actual shift in velocity (underneath any additional variability in shape and strength) rather than undergoing velocity-dependent variations that mimic a velocity shift. 

\subsection{BAL Velocity-Shift Measurements}  
\begin{figure*}
\begin{center}
\includegraphics[scale = 0.275, trim = 0 0 0 0, clip]{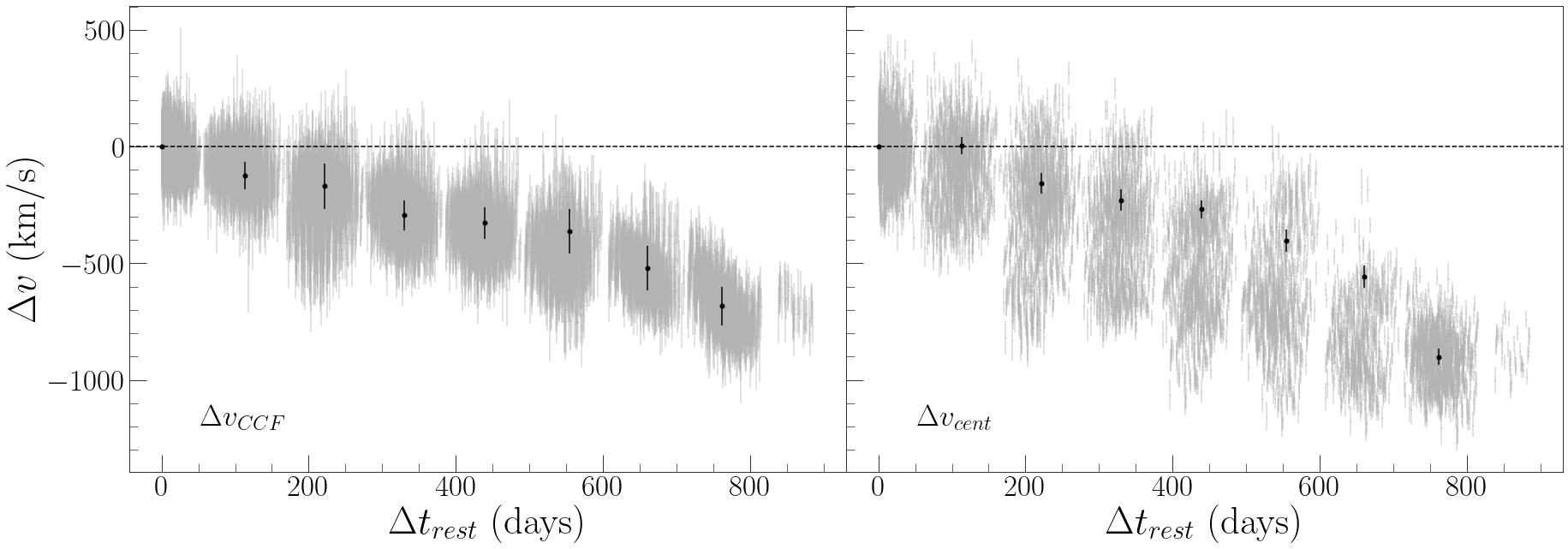} %dvccf_dvcent_vs_time.png}
\caption{The measured velocity shift as a function of time between all possible pairs of epochs, in the quasar rest frame. The left panel shows the velocity shift measured {from the CCCD ($\Delta v_{\rm CCF}$) } and the right panel shows the shift as measured by the difference in $v_{\rm cent}$ between epochs ($\Delta v_{\rm cent}$). Gray lines indicate measurements made between all individual epoch pairs; black points show measurements made between the Year 1 mean spectrum and all additional mean spectra in subsequent years.} 
\label{fig:vshift_vs_deltat} 
\end{center} 
\end{figure*}

To quantify the velocity shift in our \civ BAL, we adopt two different measurements. First, we measure the change in $v_{\rm cent}$ between spectra, $\Delta v_{\rm cent}$. This is a straightforward way to measure the change in velocity of the trough over time and quantify the acceleration of the BAL. Figure~\ref{fig:vshift_vs_deltat} shows the change in $v_{\rm cent}$ between all 8256 possible pairs of spectral epochs (each of the 129 spectra paired with every other spectrum following it). We see a clear trend of an increasing velocity shift with an increase in $\Delta t$, supporting the idea that the BAL is in fact accelerating. We do note a slight ``bifurcation" in the $\Delta v_{\rm cent}$ measurements shown in Figure~\ref{fig:vshift_vs_deltat} that arises due to different pairs of epochs that have similar $\Delta t$ (for example, measurements between day 1 and day 400, and between day 400 and day 800, will both have the same $x$-value on the plot). This suggests that the magnitude of the velocity shift changes throughout the monitoring period. 

While $v_{\rm cent}$ provides a way to characterize a velocity shift, in principle it is also sensitive to changes in the line profile. This means that velocity-dependent variability within the trough can cause a change in $v_{\rm cent}$ even if the trough itself remains within the same boundaries. While we do not see significant line-profile variability on short timescales in the BAL of RM\,613, the shape of the BAL {\it does} appear to change during the latter half of our monitoring period. By relying solely on $v_{\rm cent}$ to quantify the acceleration, we risk the possibility that our measurements are somewhat skewed by this shape change in later years. 

To help minimize the possibility that our measurements of acceleration are affected by line-profile changes, we adopt a second method of quantifying the velocity shift between two epochs: the Interpolated Cross Correlation Function (ICCF), as adapted for searches for velocity shifts in BALs by \cite{Grier16}. The ICCF procedure (e.g., \citealt{Peterson98}) was originally developed to measure time delays between light curves in reverberation mapping data -- however, the procedure can be adapted to search for velocity shifts in spectra as well. Using the CCF to measure the velocity shift may be less affected by changes in the line strength and shape than $v_{\rm cent}$ because the ICCF will measure a lower correlation coefficient when this variability is present. 

We first isolate the \civ BAL in our normalized spectra by cropping the spectra down to include only the BAL itself and about 2000~km~s$^{-1}$ of ``padding" on either side of the BAL (to allow for shifts). We implement the ICCF analysis via the {\tt PyCCF} software (\citealt{Sun18}), which works as follows: The code first measures the correlation coefficient $r$ between two spectra, and then applies a velocity shift (in increments defined by the user) and interpolates the data so that the shifted spectra lie on the same wavelength grid. The coefficient $r$ is then remeasured after applying this shift, and the spectra are shifted again --- $r$ is measured after applying all possible velocity shifts within a given range to build up the cross correlation function (CCF). The peak value of the CCF represents the velocity shift at which the two spectra are the most highly correlated. As is commonly done in reverberation-mapping studies, we measure the centroid of the CCF about the peak by including all points with values greater than 0.8$r_{\rm peak}$ to characterize the velocity shift between two spectra ($\Delta v_{\rm CCF}$). 

\begin{figure}
\begin{center} 
\includegraphics[scale = 0.3, trim = 5 32 0 0, clip]{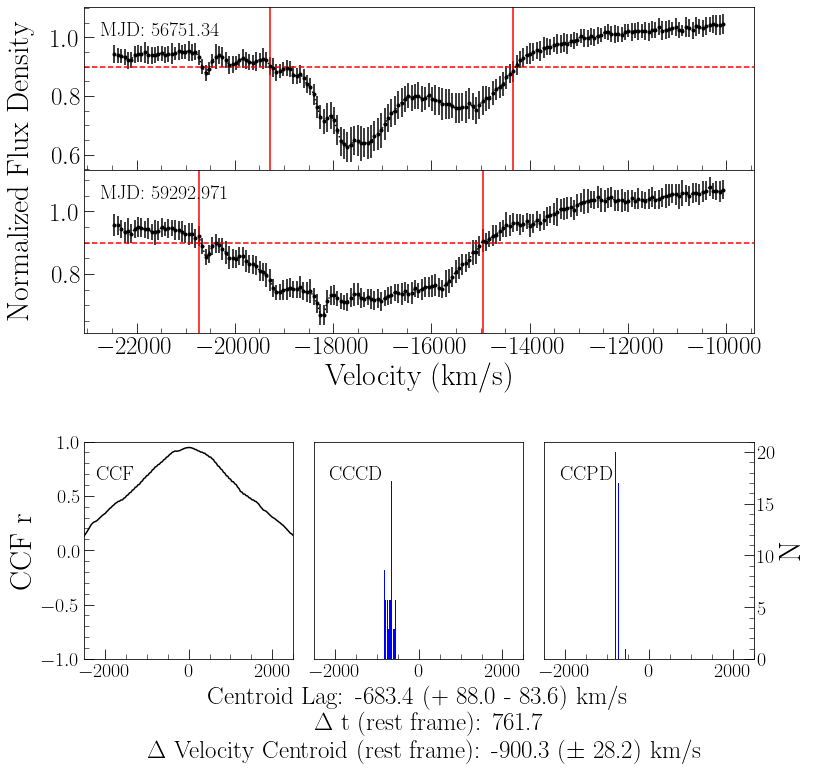} %ccf_fig_(0, 7).png}}
\caption{CCF, CCCD, CCPD analysis between the mean spectrum of 2014 (Year 1) and mean spectrum of 2021 (Year 8). The upper half of the figure shows the normalized spectra and their errors, along with the superposition of both on one figure. The vertical lines show the upper and lower boundaries of the BAL. The lower half of the figure shows the CCF, CCCD, and CCPD (see Section~\ref{sec:measurements}). } 
\label{fig:ccf} 
\end{center} 
\end{figure}

To measure the uncertainties in $\Delta v_{\rm CCF}$, we follow \cite{Peterson98, Peterson04} and employ Monte Carlo simulations as we did for the continuum fits. We perform 1,000 iterations: The spectra are altered by random Gaussian deviates scaled by their uncertainties, and the CCF is recalculated for each iteration. We then adopt the median of the cross correlation centroid distribution (CCCD) as our best velocity-shift measurement (we also keep track of the peak distribution, or CCPD), and the 1$\sigma$ uncertainties are calculated corresponding to the 68.3\% percentile of the CCCD. We generally measure uncertainties on the order of 1-2 pixels (each SDSS wavelength bin corresponds to $\sim$69 km$s^{-1}$), depending on the SNR of the individual spectra involved. Figure~\ref{fig:ccf} shows an example pair of spectra and the resulting CCF and CCCD. 

As with $\Delta v_{\rm cent}$, we measured the CCF between all 8256 pairs of spectra to explore all possible timescales. Figure~\ref{fig:vshift_vs_deltat} shows these measurements as a function of time. We again see significant velocity shifts that increase with $\Delta t$; our detection of acceleration is thus strengthened by the observations of effectively the same acceleration trends using these two independent methods. However, the CCF measured smaller shifts overall than $\Delta v_{\rm cent}$ --- we suspect that the differences are due to the changing shape of the BAL at the end of the campaign. Because $\Delta v_{\rm CCF}$ is less sensitive to these changes in the shape than $\Delta v_{\rm cent}$, it measures a smaller shift between the spectra in pairs where there has a been a substantial shape change between observations. Additionally, the ``bifurcation" seen in the $\Delta v_{\rm cent}$ measurements in Figure~\ref{fig:vshift_vs_deltat} disappears when the CCF is used, further suggesting that the increase in the velocity shifts measured by $\Delta v_{\rm cent}$ between epochs in the latter half of the observations is significantly affected by the change in shape of the BAL.

\subsection{Measurements of the Mean Spectra} 
\label{sec:meanspectra}
With $\sim$130 spectral epochs and more than 8000 possible pairs of spectra, choosing individual epochs to best quantify the acceleration across the 8 years of monitoring is somewhat challenging. To simplify things, we used the mean spectra from observations in each individual year (Figure~\ref{fig:mean_specs_by_year}), which largely have similar noise properties as one another, and made measurements of the BALs using these mean spectra for further use. We adopt the median MJD among each individual observing season as the effective MJD for each mean spectrum; this effectively gives us eight ``epochs" to work with. We then measured the EW, mean depth, and $v_{\rm cent}$ from the eight mean spectra and calculated $\Delta v_{\rm cent}$ and $\Delta v (\rm CCF)$ between sequential pairs of mean spectra. These measurements are provided in Table~\ref{Table:mean_BAL_measurements}. In addition, we provide measurements from our single HET spectrum, which was taken roughly two years after the last SDSS spectrum (see Appendix~\ref{app:het}), to characterize the activity of the BAL beyond the SDSS monitoring period. 

\begin{deluxetable*}{ccccccrrrr} 
\tablewidth{0pt} 
\tablecaption{BAL Measurements from Mean Spectra and HET Spectrum} 
\tablehead{ 
\colhead{Observing} & 
\colhead{Median} &
\colhead{$\Delta t$ (days)} & 
\colhead{EW} & 
\colhead{Mean} & 
\colhead{$v_{\rm cent}$} &  
\colhead{$\Delta v_{\rm cent}$} & 
\colhead{$a_{\rm cent}$} &
\colhead{$\Delta v(\rm CCF)$} & 
\colhead{$a_{\rm CCF}$} \\ 
\colhead{Season} & 
\colhead{MJD}  &
\colhead{(rest-frame)}  &
\colhead{(\AA)}  &
\colhead{Depth}  & 
\colhead{(km~s$^{-1}$)} & 
\colhead{(km~s$^{-1}$)} &
\colhead{(cm~s$^{-2}$)} &
\colhead{(km~s$^{-1}$)} &
\colhead{(cm~s$^{-2}$)} 
}  
\startdata 
1  &  56751.3   &  -    &  5.60$\pm$0.13   &  0.227$\pm$0.005   &  $-16807\pm$26    &  -  &  -  & - & - \\
2  &  57131.2   &  113.8    &  3.60$\pm$0.09    &  0.169$\pm$0.004   &  $-16803\pm$28    &  $4\pm$38  &  $0.04\pm$0.39 & $-135^{+45}_{-56}$ & $-1.36^{+0.46}_{-0.57}$ \\
3  &  57492.3  &  108.2    &  4.28$\pm$0.11   &  0.181$\pm$0.004   &  $-16966\pm$35    &  $-163\pm$45  &  $-1.74\pm$0.48 & $-68^{+106}_{-90}$ & $-0.72^{+1.14}_{-0.96}$ \\
4  &  57851.2   &  107.5    &  3.78$\pm$0.10   &  0.167$\pm$0.004   &  $-17036\pm$38    &  $-70\pm$51  &  $-0.75\pm$0.56 & $-105^{+69}_{-94}$ & $-1.12^{+0.75}_{-1.01}$ \\
5  &  58216.4   &  109.4    &  3.90$\pm$0.08   &  0.170$\pm$0.003   &  $-17076\pm$29    &  $-40\pm$47  &  $-0.42\pm$0.51 & $-36^{+108}_{-102}$ &  $-0.38^{+1.14}_{-1.08}$ \\
6  &  58602.3   &  115.6    &  5.00$\pm$0.11   &  0.194$\pm$0.004   &  $-17210\pm$41    &  $-134\pm$50  & $-1.34\pm$0.50 & $-5^{+83}_{-78}$ &  $-0.05^{+0.83}_{-0.78}$ \\ 
7  &  58952.9   &  105.0    &  5.41$\pm$0.14   &  0.217$\pm$0.005   &  $-17364\pm$39    & $-154\pm$57   &  $-1.70\pm$0.62 & $-186^{+106}_{-121}$ & $-2.05^{+1.16}_{-1.33}$ \\ 
8  &  59293.0   &  101.9    &  6.31$\pm$0.09   &  0.220$\pm$0.003   &  $-17708\pm$24    & $-344\pm$46  &  $-3.90\pm$0.52 & $-171^{+123}_{-103}$ & $-1.94^{+1.40}_{-1.17}$ \\ 
HET & 60094.0 &  240.0      &  3.32$\pm$0.27   & 0.13$\pm$0.01      & $-18109\pm$146   &   $-401\pm$148     &   $-1.94\pm$0.71        &     &    \\
\enddata
\tablenotetext{*}{$\Delta t$, $\Delta v_{\rm cent}$, $\Delta v(\rm CCF)$, and all acceleration measurements were all made between sequential epochs. All measurements were made using the mean spectra from each observing season, aside from the HET spectrum, which is the only available spectrum from that year.}    
\label{Table:mean_BAL_measurements} 
\end{deluxetable*}  

We also calculate the measured acceleration between each pair of mean spectra using both $\Delta v_{\rm cent}$ and $\Delta v_{\rm CCF}$. In some cases, depending on which method is used, we measure acceleration that is consistent with zero to within the 1$\sigma$ uncertainties, indicating that on one-year timescales, velocity shifts are still often too small to detect at high confidence. However, we measure acceleration with a range of values from 0.04 cm~s$^{-2}$ to 3.90 cm~s$^{-2}$ using $\Delta v_{\rm cent}$ (hereafter we will refer to the acceleration calculated from $\Delta v_{\rm cent}$ as $a_{\rm cent}$), and ranging from 0.05 cm~s$^{-2}$ to 2.05 cm~s$^{-2}$ when using $\Delta v_{\rm CCF}$ to calculate the acceleration (hereafter, $a_{\rm CCF}$). We suspect that some of the differences between $a_{\rm cent}$ and $a_{\rm CCF}$ measurements are a result of the shape change that occurred in the final few years of monitoring; we thus see a more significant increase in $a_{\rm cent}$ over that period, but suspect that not all of the measured velocity shift can be attributed to the actual acceleration of the gas. Previous studies (e.g., \citealt{Grier16}, \citealt{Byun22}, \citealt{Xu20}) that report possible acceleration candidates have also measured a wide range of acceleration values (ranging from $-0.25$ to 1.5~cm~s$^{-2}$), supporting the idea that the acceleration can be variable in magnitude.  

\begin{figure}
\begin{center}
\includegraphics[scale = 0.3, trim = 0 10 0 0, clip]{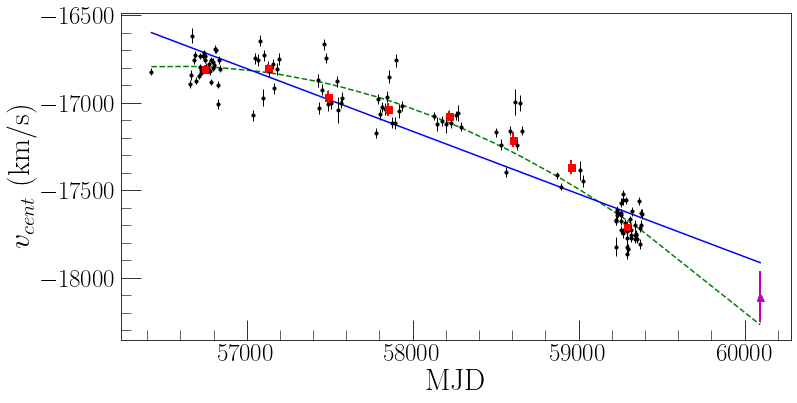} %vcent_fits.png}
\caption{The velocity centroid $v_{\rm cent}$ as a function of time. Black points represent measurements made from each individual SDSS spectrum and the magenta triangle represents the measurements made from the HET spectrum. Red squares show the values measured from the mean spectra of each individual year (see Section~\ref{sec:meanspectra}). A linear fit to the data is shown by the blue line, and a 2nd order polynomial fit is shown as a dashed green line. } 
\label{fig:vcent_linefit} 
\end{center} 
\end{figure} 

Examination of Figure~\ref{fig:things_vs_mjd} indicates a slow decrease in $v_{\rm cent}$ over the monitoring period; however, the relationship between $v_{\rm cent}$ and time does not appear to be linear. Figure~\ref{fig:vcent_linefit} again shows $v_{\rm cent}$ as a function of time, but this time we have fit a linear relation to the data and we see that the trend is not well fit (the reduced $\chi^2$ of a linear fit to the data is 20.9) -- there is an overall curvature to the trend, particularly during the latter half of the campaign. We instead fit a 2nd order polynomial to the data and we find this fit in much better agreement than the linear fit (the reduced $\chi^2$ of the 2nd order polynomial is 10.2), though both are poor fits in general, as there is significant short-timescale variability on top of the long-term trend. This would suggest that the magnitude of the acceleration is increasing over time. However, this interpretation is complicated by the fact that we also see a substantial change in the shape of the BAL during the latter half of the campaign (e.g., Figure~\ref{fig:mean_specs_by_year}). For the first half of the campaign, $v_{\rm cent}$ and $v_{\rm CCF}$ measure changes that are roughly on par with one another; however, in the latter half of the campaign, where the shape changes become more substantial, $v_{\rm CCF}$ measures overall smaller shifts than $v_{\rm cent}$. This suggests that the non-linearity in the relationship between $v_{\rm cent}$ and time is due to the changing shape of the BAL over this time rather than actual acceleration of the gas. 

In addition to measuring $a$ between sequential years, we also measure the acceleration between Years 1 and 8 to obtain the average acceleration over the entire monitoring period. {We measure $\Delta v_{\rm cent}$~=~$-901~\pm~35~$km~s$^{-1}$ and $\Delta v_{\rm CCF}$~=~$-683^{+88}_{-84}$~km~s$^{-1}$, corresponding to {$a_{\rm cent}$~=~$-1.37\pm$0.05~cm~s$^{-2}$} and {$a_{\rm CCF}$~=~$-1.04^{+0.14}_{-0.13}$ cm~s$^{-2}$.}} These values are the same order of magnitude as measurements from previous work, which examine BALs in quasars that have broadly similar properties. 

\subsection{Evolution of Trough A ``Sub-Troughs"}  
\label{sec:subtroughs}
\begin{figure}
\begin{center}
\includegraphics[scale = 0.3, trim = 0 0 0 0, clip]{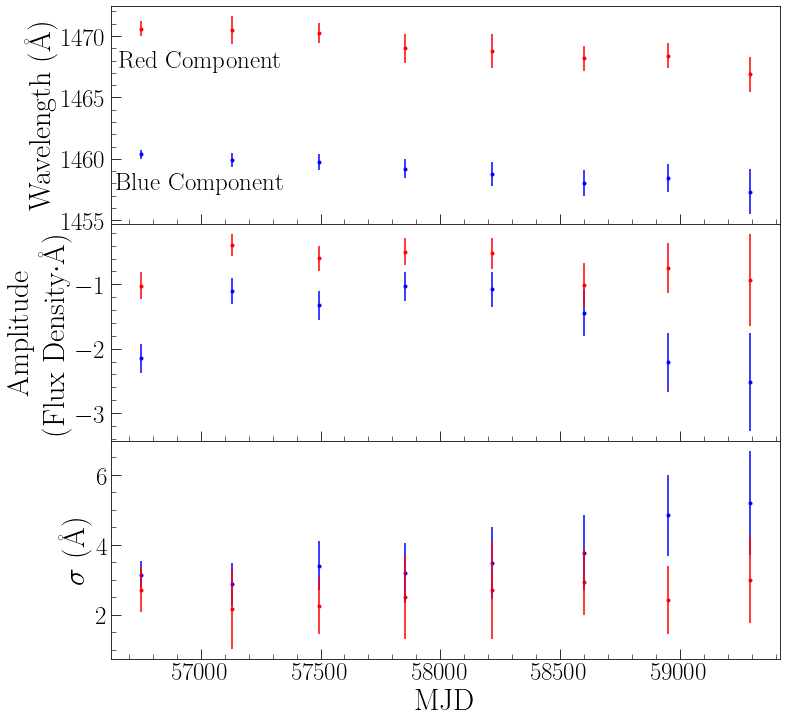} % gauss_3things.png}

\caption{The model fit parameters of the two Gaussian components of the Trough A model as a function of time. Blue symbols represent the higher-velocity, or bluer, Gaussian sub-trough component, and red represents the lower-velocity, or redder sub-trough component. The top panel shows the center wavelength of each component; the middle panel shows the amplitude of the Gaussian fits, and the bottom panel shows the width $\sigma$ of the fit. The units of the amplitude are normalized flux density times \AA.} 
% the real units are Normalized Flux Density * \AA
\label{fig:gauss_center_by_year} 
\end{center} 
\end{figure}

Figures~\ref{fig:trailed_spec} and \ref{fig:mean_specs_by_year} clearly indicate a change in overall shape of Trough A; at the beginning of the monitoring period, we see two separate ``sub-troughs" within the BAL, whereas the last year of monitoring shows a more broad, single-component trough. In an attempt to disentangle  the behavior of these individual sub-troughs within Trough~A from the observed velocity shift of the BAL, we model the BAL in each mean spectrum using two Gaussian profiles. We include only pixels that fall within the formal BAL limits in each mean spectrum --- i.e., our fits are restricted to the region where the normalized flux density lies below 0.9 (see Appendix~\ref{app:modelfits} to see the actual model fits). We then examine the behavior of the two Gaussian components over time. Figure~\ref{fig:gauss_center_by_year} shows the measured parameters of the Gaussian component representing each sub-trough (hereafter referred to as the red sub-trough and blue sub-trough) as a function of time. 

We see that {\it both} sub-troughs are shifting blueward, though at slightly different rates: The red sub-trough begins with a line center of roughly 1470.5\,\AA\ and shifts blueward to a line center of 1467\,\AA\ (a shift of roughly 3.5\,\AA\ over the observing period), and the blue sub-trough begins with a line center of roughly 1460.5\,\AA\ and shifts to 1457.5\,\AA, corresponding to a total shift of roughly 3\,\AA. The amplitude of the Gaussian fit of the blue sub-trough follows the overall EW trend during this period, starting off strong, followed by a significant decrease in amplitude in Years 2-6 and then a substantial increase in Years 7 and 8. The amplitude of the Gaussian representing the red sub-trough remains substantially weaker than that of the blue sub-trough throughout the campaign, and does not increase in amplitude during Years 7 and 8; by Year 7, the model is best fit by one major Gaussian (the blue sub-trough) that has significantly widened, with the second Gaussian contributing minimally. The measured widths of the blue Gaussian component increases over the monitoring period, whereas the width of the red component remains lower throughout. 

The behavior of the two Gaussian components in our model fits suggests that both of the sub-troughs are shifting; however, the right sub-trough seems to be shifting blueward at a slightly higher rate than the left sub-trough: Between Years 1 and 8, the center of the red component shifted in velocity by 754 km~s$^{-1}$ (from $-15405$ km~s$^{-1}$ to $-16159$ km~s$^{-1}$), whereas the blue component moved by 625 km~s$^{-1}$ (from $-17485$ km~s$^{-1}$ to $-18110$ km~s$^{-1}$). This, combined with the increase in EW of the blue sub-trough of the BAL during the last two years of monitoring, causes the two sub-troughs to appear to merge into a single, broader trough while they shift. However, our HET spectrum taken two years after the SDSS observations indicates that the two components still remain distinct from one another, suggesting that they continue to behave as separate entities. 

\subsection{Measurements of \civ Troughs B and C} 
In addition to our measurements of Trough A, we also investigate the two narrower \civ absorption features present in RM\,613, Troughs B and C (see Figure~\ref{fig:mean_normalized}). Trough B has a velocity width of $\sim1311$~km~s$^{-1}$, formally classifying it as a ``mini-BAL" (e.g., \citealt{Hall13}), and a centroid velocity of about $-4800$~km~s$^{-1}$.  Trough C is also formally classified as a mini-BAL, with a velocity width of $\sim1380$~km~s$^{-1}$ and a centroid velocity of about $-1000$~km~s$^{-1}$. \cite{Grier15} examined the behavior of both of these troughs during the first year of SDSS-RM monitoring and found that Trough B showed similar behavior in strength (EW and mean depth) as BAL Trough A, but Trough C did not show any significant variability in strength. The coordination between Troughs A and B implied that the observed variability is due to a change in ionization state of the gas. 

\begin{figure}
\begin{center}
\includegraphics[scale = 0.1, trim = 0 0 0 0, clip]{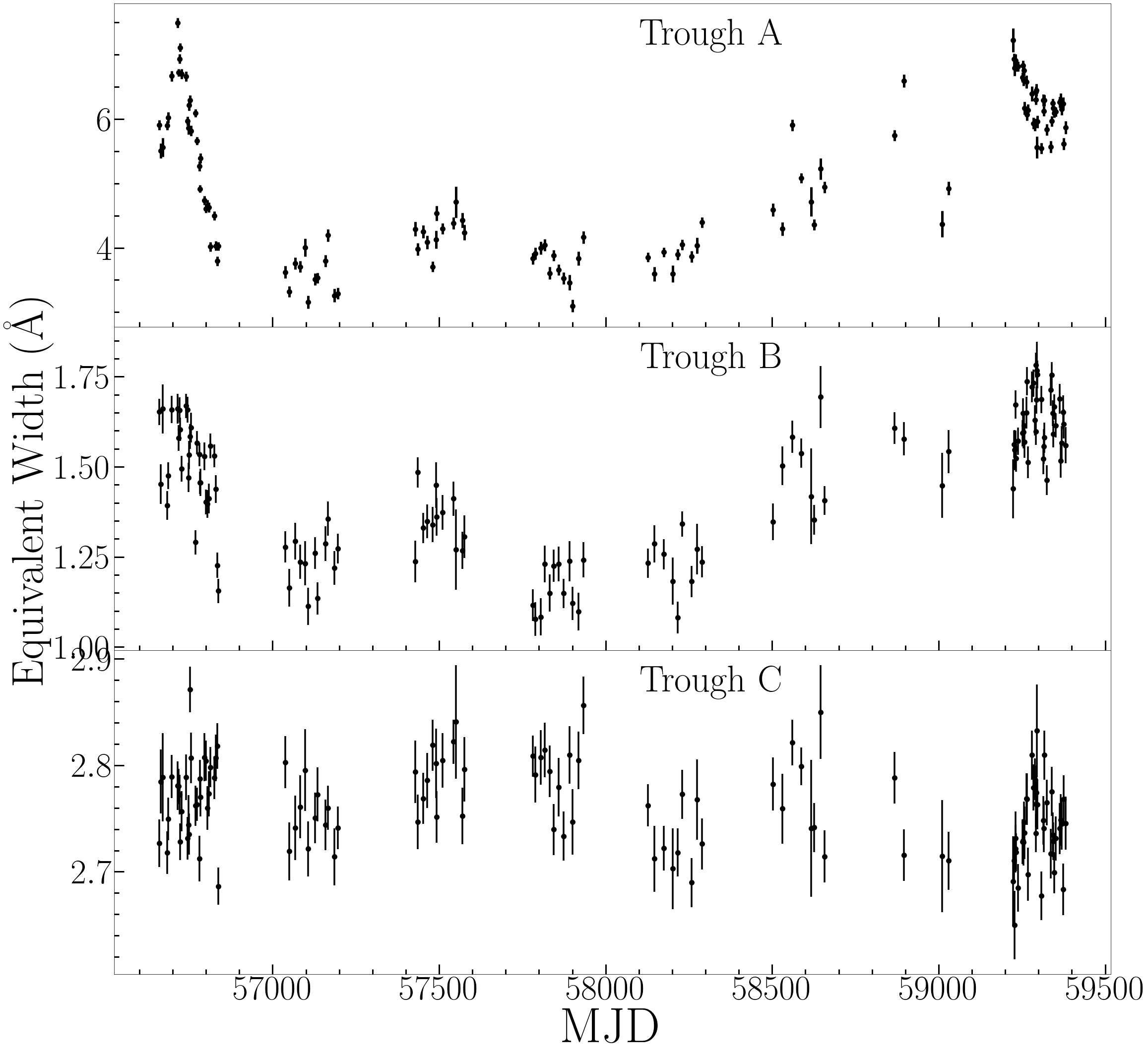} %ew_3_troughs.png}
\caption{The rest-frame EW of Troughs A, B, and C as a function of time (MJD).} 
\label{fig:troughs_b_c} 
\end{center} 
\end{figure} 

Figure~\ref{fig:troughs_b_c} shows the EW as a function of time for all three troughs for the eight-year monitoring period. We again find that the EW and depth of Trough B tracks that of Trough A; while there is somewhat more noise (likely due to the BAL being superimposed on the also-variable \civ emission line), we see similar behavior in the strength of the BAL. However, we do not see any velocity shift in Trough B; the upper and lower boundaries of Trough B remain constant to within the uncertainties during the entire monitoring period. Trough C does not show any of the same variability trends in strength as the other two troughs -- the EW shows very little variability beyond statistical scatter. Similar to Trough B, Trough C remains stable in velocity throughout the duration of our observations. 

\section{Discussion} 
\label{sec:discussion} 
While there have been other reports of monolithic velocity shifts in BALs over the years (see Section~\ref{introduction} for a list of such reports), this is the first time we have made a detection in a source with such dense time sampling to allow us to track the behavior of the BAL on short-to-long timescales. This allows us to evaluate several different mechanisms for producing the observed velocity shifts as well as to evaluate the efficacy of previous searches for BAL velocity shifts/acceleration. 

\subsection{Possible Causes of Velocity Shifts} 
We below consider a few possible models to describe the behavior observed in RM\,613 and evaluate the suitability of these models in producing this behavior. 

\subsubsection{Changes in ionization state} 
We first consider the possibility that changes in the ionization state of the gas or the covering factor have caused the observed velocity shift. While this possibility is difficult to rule out in previous cases where we had only 1-2 spectra to examine, the sheer number of observations we have of this source suggests that this is unlikely in the case of RM\,613. It would take a remarkable coincidence of circumstances for the BAL to slowly and consistently vary in just such a way as to appear to show a monolithic increase in velocity over such a long period of time. For example, our modelling of Trough A as two Gaussians might have yielded two Gaussians at fixed velocities, with the red one weakening and the blue one strengthening to give the impression of acceleration. Instead, the best-fit model found increasingly negative velocities for both Gaussians.

\begin{figure}
\begin{center}
\includegraphics[scale = 0.30, trim = 20 0 80 0]{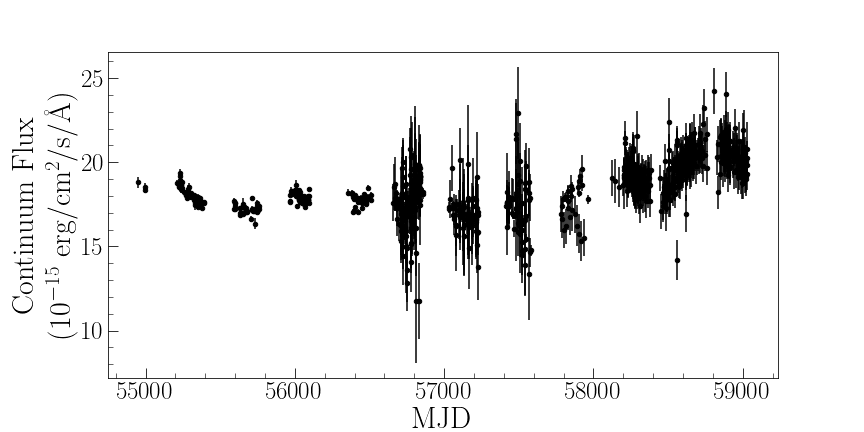} %conflux_mjd.png
\caption{Continuum flux, as measured from both photometric and spectroscopic observations of the quasar over time (light curves are from \citealt{Shen23}).} 
\label{fig:con_flux} 
\end{center} 
\end{figure}

In addition, inspection of the continuum light curve over the monitoring period (Figure~\ref{fig:con_flux}; \citealt{Shen23}) shows no unusual variability in the continuum flux of RM\,613 during this time. However, the continuum flux is slightly elevated in Years 7 and 8 of the campaign, which also corresponds with the period in which we see the most dramatic changes in shape and the highest rates of acceleration. We examined the light curves of both the \civ and \heii\ 1640\,\AA\ emission lines, which sometimes trace the ultraviolet continuum more closely than the optical continuum (e.g., \citealt{Derosa15}), and we similarly see no unusual or extreme variability in either. The slope of the continuum increases in the negative direction as the quasar gets brighter (i.e., as the quasar continuum luminosity increases, the quasar gets bluer), again showing no unusual behavior throughout the monitoring period. 

\subsubsection{Geometric Effects}
\label{sec:discussion_geo}
We next consider the possibility that geometric effects may produce the observed velocity shift; for example, if the outflow is launched from a rotating disk, its continued rotation may also cause an observed velocity shift (e.g., \citealt{Hall13}, \citealt{Grier16}) even if the speed of the outflow remains constant. However, assuming the rotating wind is silhouetted against the accretion disk, we would only see the wind that is appearing to {\it decelerate} rather than accelerating: the wind rotates into view with its maximum blueshift, gradually shifts to moving perpendicular to our line of sight (no shift), then becomes redshifted until it rotates out of view with its maximum redshift. Thus the rotation of a wind silhouetted against the accretion disk cannot produce our observed increase in velocity. 

There is another possible geometric effect that may cause a velocity shift: a change in flow angle at constant speed. If the velocity change is due to a change in flow direction, for example due to a change in the azimuthal and/or polar angle of an outflow from an accretion disk where the outflow crosses our link of sight,
we can determine the change in angle required to produce the observed velocity shift as follows. If $\theta_{\rm LOS}(t)$ is the angle between the flow and the line of sight and $|v_{\rm true}|$ is the speed of the outflow, then:
\begin{equation} 
 v_{\rm LOS}(t) = v_{\rm true} \cos \theta_{\rm LOS}(t). 
\end{equation}
With $t = t_1$ for the first observing season, we can solve for $v_{\rm true}$ and obtain the line of sight velocity at any other time $t$ when the outflow has the same $v_{\rm true}$ but makes an angle $\theta_{LOS}(t)$ with our line of sight:
\begin{equation} 
v_{LOS}(t) = v_{\rm LOS}(t_1) \cos \theta_{\rm LOS}(t) / \cos \theta_{\rm LOS}(t_1).
\end{equation}

With our measured $v_{\rm LOS}(t_1)=16800$~km~s$^{-1}$, the minimum value of $\theta_{\rm LOS}(t_1)$ is 19 degrees (which requires $\theta_{\rm LOS}(t_8)=0$ degrees in the 8th observing season). If $\theta_{\rm LOS}(t_1)=45$ degrees, $\theta_{\rm LOS}(t_8)=41.8$ degrees. The maximum plausible $\theta_{\rm LOS}(t_1)=73.7$, as that yields $v_{\rm true}~=~60,000$~km~s$^{-1}$, matching the fastest known UV outflows (\citealt{Rogerson15}); in that case, $\theta_{\rm LOS}(t_8)=72.8$ degrees. Thus, depending on the initial angle of the flow to the line of sight, a variation from 1 -- 19 degrees in that angle can explain the observed acceleration as a change in flow direction at fixed speed. This model cannot be ruled out, but does not explain all of the observed changes in $v_{\rm cent}$, which shows variations that are poorly described by a linear change (Figure~\ref{fig:vcent_linefit}). It is possible, however, that this effect, combined with the overall shape change that occurs in the BAL toward the latter half of the observations, may explain our observations. 

\subsubsection{Gas Dynamics} 
Another possible scenario is that the gas may be accelerating due to hydrodynamical effects such as overpressure, buoyancy, or entraining\footnote{{We note that an increase in magnetohydrodynamic driving is also possible if the outflow is strongly magnetized, either by magnetic field rearrangement or by other effects (e.g., \citealt{Granot11})}}.  For example, \cite{Waters21} discuss how bubbles of hot gas can form in a disk atmosphere and rise out of it, accelerating some cooler gas out of the atmosphere in the process. If the absorbing gas structure (traveling at $\sim$17000 km~s$^{-1}$) moves into a region of lower pressure, the structure will expand. Such expansion would only happen at the sound velocity ($\sim$10 km~s$^{-1}$) or, at most, the turbulent velocity. A very large turbulent velocity of 900~km~s$^{-1}$ would be required to produce the observed velocity shift in our observations, which seems unlikely. This model would also predict that the acceleration is temporary, lasting only until the structure comes into pressure equilibrium again. Future observations of RM\,613 will be revealing in this regard.  

Alternatively, if the absorbing structure moves into a region where the surrounding gas is denser than the absorbing gas, the absorbing gas will accelerate away from the quasar due to a buoyancy force.  However, denser surrounding gas should be lower-ionization than the absorbing gas and should itself produce absorption, so a pure buoyancy scenario seems implausible.

The visualizations of simulated disk wind outflows presented by \cite{Proga12} illustrate how geometric and hydrodynamical effects can produce velocity shifts.  In their Figure 1, between panels 3, 4 and 5 the blue edge of an absorption trough increases in outward velocity due to the motion of gas of different outward velocities across accretion-disk regions of different surface brightnesses, mimicking acceleration of a single flow structure. However, the red edge of the simulated trough does not exhibit a significant shift, which may indicate that a case like ours with a shift in velocity at both trough edges is more likely to be acceleration by radiation pressure, as discussed below.

\subsubsection{Acceleration due to Radiation Pressure} 
\label{sec:discussion_rad}
We next consider that the observed acceleration could be due to an increase in speed of the outflow by incident ionizing radiation (e.g., \citealt{Murray95}). 
This could be a case of our line of sight intersecting an outflow in a location where it is being accelerated to its terminal outflow velocity $v_\infty$, or a case of a previously coasting outflow being newly accelerated.

In \S\,4.1 of \citet{Grier16}, we provided an equation for the gas velocity and acceleration as a function of distance $r$ from a black hole for a \citet{Murray95} disk wind launched from a radius $r=r_L$.
%The acceleration in such a wind starts at zero at $r=r_L$, rises steeply to a maximum at $r=1.65r_L$, and then decreases roughly as $r^{-2}$.   
The velocity is given by $v(r)=v_\infty(1-r_L/r)^{1.15}$ and the ratio of the acceleration to the velocity satisfies $a(r)/v(r)=1.15v(r)r_L/r^2$.
If we adopt an average acceleration of $a=1.21\pm 0.12$~cm~s$^{-2}$ at a velocity of
$v=1.6807\times 10^9$~cm~s$^{-1}$ and assume a plausible value for $r_L$, we can find values of $r$ and $v_\infty$ for which this model will match the observations.
Following \citet{Grier16} and assuming $M_{BH}\sim 2\times 10^9 ~M_\odot$ and $r_L=3.6\times 10^{17}$~cm = 0.12 pc = 610 $R_{\rm Sch}$, we find $r=2.73r_L=9.83\times 10^{17}$~cm = 0.33 pc, and $v_\infty=28,400$~km~s$^{-1}$.
These numbers are only approximate, but they do indicate that to explain our observations with this model of gas accelerating to a fixed $v_\infty$, the gas needs to be located within a parsec of the central engine. For example, doubling $r$ requires quadrupling $r_L$ (so that $r=1.365r_L$) and more than doubling $v_\infty$ to an implausible $76,600$~km~s$^{-1}$, and cutting $r_L$ in half yields $r=3.86r_L$, or 0.23 pc, and $v_\infty=23,700$~km~s$^{-1}$.

If we consider that the acceleration may have increased over time (Figure \ref{fig:vcent_linefit}), then in this model either the gas is at a radius $r<1.65r_L$ (since this model's acceleration is only increasing at such small radii) or the terminal velocity of the wind is changing with time. The terminal velocity is approximately $v_\infty \simeq v_K(r_L)\sqrt{F_M \Upsilon_{\rm eff}}$ \citep{Laor02}.
Here, $v_K(r_L)$ is the Keplerian circular speed at the launch radius and the effective Eddington ratio $\Upsilon_{\rm eff}=L_{\rm inc}/L_{\rm Edd}$ is the ratio of the quasar's luminosity incident on the wind gas (after accounting for any absorption interior to the wind) to its Eddington luminosity, and $F_M$ is the force multiplier acting on the wind gas. 
As discussed in \S\,2 of \citet{Murray95}, the force multiplier is larger where $\frac{\partial v_r}{\partial r}$ is large (i.e., in regions of large acceleration) and where the ionization is low.

\citet{Naddaf23} presents a different model in which radiation pressure on dust is the primary driver for the BAL outflow. From Fig.\ 4 of that paper, the launch radius for their model wind is about 10-20 times larger than for our model above, ranging from $(6-1.2)\times 10^3 R_{\rm Sch}$. Dust could provide this wind with a larger $F_M$, enabling it to reach the observed outflow velocities from a larger $r_L$. This wind would have a similar $v(r)$ profile to our model, with a length scale for reaching terminal velocity similar to the launch radius. Thus, in this dust-driven wind model, the BAL gas observed to be accelerating in this quasar would need to be located within 10-20 pc of the central engine.

As mentioned at the start of this section, it is also possible that the gas was previously coasting but has recently begun to accelerate.
While we do not observe any dramatic/extreme variability in the optical continuum flux, small increases in the optical continuum flux may trace a larger increase in shorter-wavelength flux (such as that in the extreme ultraviolet) which could accelerate the outflow gas without over-ionizing it.
A simple increase in the ionizing continuum level seems unlikely to always result in acceleration, since continuum variability is seen in BAL quasars without acceleration and deceleration being common (e.g., \citealt{Filizak13, Grier16}). An unusually large or sustained ionizing continuum increase still might explain the acceleration, and we note that the  velocity change in RM\,613 was observed to be largest after MJD 58500, when the near-UV continuum flux and the EWs of Troughs A and B were all increasing.
However, \citet{wywf} show that increases in the near-UV continuum are accompanied by increases in BAL trough EWs only $\sim$25\% of the time (their Figure 7b), from which they conclude that most carbon atoms in the gas producing BAL troughs are in states more highly ionized than C\,{\sc iv}.

In summary, it could be that the gas in the BAL trough observed in this object is located at a relatively small radius where it is still being accelerated by the quasar's ultraviolet radiation ($\lesssim 1$ pc in the \citealt{Murray95} model; $\lesssim 10-20$ pc in the \citealt{Naddaf23} model), even absent any increase in that radiation field.
Alternatively, if the gas was previously coasting, we suggest that an unusual change in the ionizing SED consisting of an increase in ionizing flux and/or a softening of the ionizing spectrum could in principle explain both the acceleration and the increase in the \civ absorption EW, by producing more photons that scatter off and accelerate ions in the BAL gas but fewer photons that ionize C\,{\sc iv}.
We leave detailed study of these possibilities to future work.

\subsection{On the Efficacy of Searches for BAL Acceleration} 
\label{sec:discussion_eff}
We are in the unique position of having more than a hundred epochs of spectra for our source; this makes the identification of acceleration fairly straightforward, as we can rule out velocity-dependent variability that mimics a velocity shift simply by visual inspection. However, in most cases, reports of velocity shifts/acceleration in BAL quasars have only a handful of epochs, frequently separated by several years. To date there has only been one systematic search for BAL acceleration in a large sample of quasars (\citealt{Grier16}; hereafter G16), and this study included sources with only three epochs of observations. G16 provide a suggested procedure for identifying and quantifying the acceleration of a BAL based on the cross-correlation of spectra from two epochs and also requiring that the overall shape of the BAL remain constant between the observed epochs. This second criterion was included by necessity --- with only 2-3 epochs, it is plausible that the cause of observed velocity shifts is changes in the ionization state of the gas or the transverse motion of gas into the line of sight, rather than the actual acceleration of the outflow. 

G16 report a detection rate of acceleration in $\sim1\%-10\%$ of quasars in their study. However, this may not be the true incidence of acceleration in BALs. Due to observational constraints, such as the limited number of spectra with which most acceleration candidates are identified, G16 may not have been able to identify all cases in which the BAL actually accelerated. With $\sim$130 spectra providing us with a solid detection in RM\,613, we can estimate, via simulations, the likelihood that we would have detected the acceleration in RM\,613 if we only had 3 epochs of spectra over this time period.

As a way of determining this likelihood, we perform a series of simulations using our data from RM\,613. We divide the spectra into three groups: Group 1 contains spectra from years 1-3, Group 2 spans years 4-6, and Group 3 spans years 7 and 8. We then randomly draw a spectrum from each set to produce a set of three epochs; division into the three groups is meant to enforce a time separation between spectra that is on par with that between spectra used by G16. We then carry out the tests defined by G16 to determine whether or not those three epochs are indicative of acceleration: We first measure the CCF between spectra 1 and 2, and between spectra 2 and 3, to determine if a velocity shift is present between these pairs. We then perform a $\chi^2$ test between these pairs both before and after having applied the measured velocity shift from the CCF. If the observed velocity shift is more than 3$\sigma$ inconsistent with zero and the original spectra are a ``bad" match, but the shifted spectra are a ``good" match (as determined by the $p$-value measured in the $\chi^2$ test), the pair of spectra is considered to have passed the criteria for acceleration. If both pairs (epochs 1-2 and 2-3) pass these tests, G16 determined that they are likely to show real acceleration signatures and are considered ``acceleration candidates". 

We perform this test with our RM\,613 data 1000 times, randomly drawing epochs within each group and carrying out the G16 test for acceleration between the three randomly selected spectra. The results were quite striking: Out of 1000 draws, only 1 produced an ``acceleration candidate" that passed the tests defined by G16. This low success rate has two potential causes: First, there is a much shorter timescale between the first pair of observations used in our study compared to those of G16 (the median $\Delta t_{\rm rest}$ between epochs 1 and 2 in our random draws was 322 days in the quasar rest frame, whereas the median $\Delta t_{\rm rest}$ between epochs 1 and 2 in the quasar sample examined by G16 is 1146 days).\footnote{We note, however, that the distribution of $\Delta t_{\rm rest}$ between epochs 2 and 3 was on par with that of the sample of G16.} The shorter timescales involved mean that the velocity shifts will be smaller, so we more rarely detect the velocity shift between epochs 1 and 2 at $> 3\sigma$ significance in our simulations simply because there was not enough time for a large velocity shift to occur. Secondly, as noted above, the strength of the BAL in RM\,613 is quite variable throughout the campaign and there is significant shape variability at the end of the monitoring period, causing epochs that are well separated in time to fail the required $\chi^2$ tests. 

Based on these simulations, we conclude that studies searching for acceleration that only include a handful of epochs, such as that of G16, will miss ``real" acceleration much of the time, particularly if those epochs are not separated in time by at least a few years in the quasar rest frame. The actual rate of acceleration in BALs is likely much higher than that reported by G16 --- the data, however, have thus far been generally insufficient for detection. High-cadence studies of BAL quasars over long timescales will be required to determine the actual rate of acceleration. 

\section{Summary}
\label{summary}
Over the years, roughly a dozen cases of significant velocity shifts in BALs have been reported (see Section~\ref{introduction} and references therein). In those cases, only a handful of spectra were used, causing difficulties with disentangling line-profile variability with actual shifts in velocity. We have investigated the variability of a high-velocity \civ BAL trough with $\sim$130 spectroscopic observations spanning more than eight years, allowing us to explore variability and acceleration on both short and long timescales. Our major findings are as follows:  

\begin{enumerate} 
%\item We observe the high-velocity \civ BAL Trough~A to vary in EW, mean depth, and $v_{\rm cent}$ throughout the campaign. As noted in previous studies, Trough~A shows significant variability in its EW on timescales down to a few days in the quasar rest frame (Figure~\ref{fig:things_vs_mjd}). 
\item We have observed a long-term velocity shift in the \civ BAL of the quasar RM\,613 over the monitoring period; the density of our observations indicates that this observed shift is not due solely to velocity-dependent variability in the BAL profile that might mimic an acceleration signature (Figures~\ref{fig:trailed_spec} and \ref{fig:mean_specs_by_year}). 
\item We observe a very slow change in the shape of the BAL that becomes noticeable in the second half of the observing period. We do not see significant changes in shape on shorter timescales; however, over the span of our observations, the BAL changed from a two-pronged feature to a single smooth trough (Figures~\ref{fig:trailed_spec} and \ref{fig:mean_specs_by_year}). 
\item Combining all spectra taken within an individual observing season, we measure the acceleration of the trough between each observing season (Table~\ref{Table:mean_BAL_measurements}), as well as over the entire monitoring period. Using the CCF, we measure the average velocity shift over the eight-year period to be {$-683^{+89}_{-84}$~km~s$^{-1}$, which corresponds to an average acceleration of $a$~=~$-1.04^{+0.14}_{-0.13}$~cm~s$^{-2}$ }. The magnitude of this acceleration varies throughout the monitoring period (Figures~\ref{fig:vshift_vs_deltat} and \ref{fig:vcent_linefit}).  
\item We consider a variety of possible causes of the observed velocity shift and determine that our observations are most likely to be explained by geometric effects (Section~\ref{sec:discussion_geo}) or the acceleration of the outflow by radiation pressure (Section~\ref{sec:discussion_rad}). If our line of sight intersects the outflow in its acceleration region, that places the observed gas within 1-20 pc of the black hole, depending on the wind model adopted.
\item Our simulations suggest that studies such as G16 that search for velocity shifts in small samples are likely to miss actual acceleration much of the time due to short time baselines and the prevalence of variability in the shape and strength of BALs (Section \ref{sec:discussion_eff}).
\end{enumerate}  

Observations of BAL velocity shifts, the implied cause of which is the acceleration of outflow material, are difficult for a myriad of reasons. We used this unique dataset to investigate both the the short and long-term variability of this particular BAL {\it simultaneously}, which allowed us to determine that this BAL showed both line-profile variability and a distinct shift in velocity. Future studies focused on BAL acceleration will require long time-baseline, high-cadence observations to determine the frequency of such behavior as well as to disentangle signatures of line-profile variability from velocity shifts due to acceleration of the outflow. The SDSS-V BHM program will include observations on BAL quasars on several different timescales. The multi-epoch spectroscopy part of BHM aims to obtain a handful of observations of hundreds to thousands of quasars, though with a cadence similar to that from G16 we expect the yield of BAL-acceleration candidates from that sample to be low. However, we expect that there are roughly 200 BAL quasars in the sample that is currently being observed by the BHM-RM program, which will have a high observational cadence similar to that in this study --- these data will be particularly instructive in searches for BAL acceleration. Additional surveys using multi-object spectrographs will aid further in the construction of a statistically significant sample of BAL acceleration candidates. \\

%\begin{acknowledgments}
RW and CJG acknowledge support from NSF grants AST-2310211 and AST-2309930. Support for this research was provided by the Office of the Vice Chancellor for Research and Graduate Education at the University of Wisconsin–Madison with funding from the Wisconsin Alumni Research Foundation. PBH acknowledges support from NSERC grant 2023-05068. WNB acknowledges support from NSF grant AST-2106990. YS acknowledges partial support from NSF grants AST-2009947 and AST-2108162. Y.H. was supported as an Eberly Research Fellow by the Eberly College of Science at the Pennsylvania State University. CR acknowledges support from the Fondecyt Iniciacion grant 11190831 and ANID BASAL project FB210003. RJA was supported by FONDECYT grant number 123171 and by the ANID BASAL project FB210003. MLM-A acknowledges financial support from Millenium Nucleus NCN19-058 and NCN2023${\_}$002 (TITANs). C.A.N. thanks support from CONAHCyT project Paradigmas y Controversias de la Ciencia 2022-320020 and DGAPA-UNAM grant IN111422.

Funding for the Sloan Digital Sky Survey V has been provided by the Alfred P. Sloan Foundation, the Heising-Simons Foundation, the National Science Foundation, and the Participating Institutions. SDSS acknowledges support and resources from the Center for High-Performance Computing at the University of Utah. The SDSS web site is \url{www.sdss.org}.

SDSS is managed by the Astrophysical Research Consortium for the Participating Institutions of the SDSS Collaboration, including the Carnegie Institution for Science, Chilean National Time Allocation Committee (CNTAC) ratified researchers, the Gotham Participation Group, Harvard University, Heidelberg University, The Johns Hopkins University, L'Ecole polytechnique f\'{e}d\'{e}rale de Lausanne (EPFL), Leibniz-Institut f\"{u}r Astrophysik Potsdam (AIP), Max-Planck-Institut f\"{u}r Astronomie (MPIA Heidelberg), Max-Planck-Institut f\"{u}r Extraterrestrische Physik (MPE), Nanjing University, National Astronomical Observatories of China (NAOC), New Mexico State University, The Ohio State University, Pennsylvania State University, Smithsonian Astrophysical Observatory, Space Telescope Science Institute (STScI), the Stellar Astrophysics Participation Group, Universidad Nacional Aut\'{o}noma de M\'{e}xico, University of Arizona, University of Colorado Boulder, University of Illinois at Urbana-Champaign, University of Toronto, University of Utah, University of Virginia, Yale University, and Yunnan University. 

This work includes observations obtained with the Hobby-Eberly Telescope (HET), which is a joint project of the University of Texas at Austin, the Pennsylvania State University, Ludwig-Maximillians-Universitaet Muenchen, and Georg-August Universitaet Goettingen. The HET is named in honor of its principal benefactors, William P. Hobby and Robert E. Eberly. The Low Resolution Spectrograph 2 (LRS2) was developed and funded by the University of Texas at Austin McDonald Observatory and Department of Astronomy, and by Pennsylvania State University. We thank the Leibniz-Institut fur Astrophysik Potsdam (AIP) and the Institut fur Astrophysik Goettingen (IAG) for their contributions to the construction of the integral field units. We acknowledge the Texas Advanced Computing Center (TACC) at The University of Texas at Austin for providing high performance computing, visualization, and storage resources that have contributed to the results reported within this paper. \\
%\end{acknowledgments}

\appendix  

\section{HET Spectrum Analysis} 
\label{app:het}
\begin{figure}
\begin{center}
\includegraphics[scale = 0.4, trim = 0 0 0 0, clip]{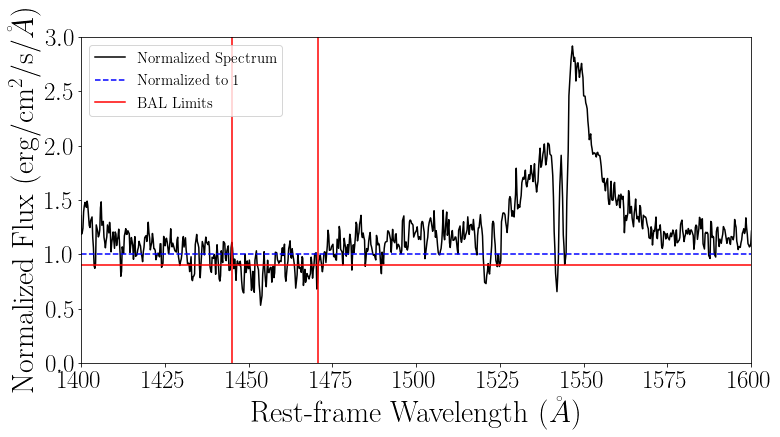} %het_bal.png
\caption{The HET spectrum (black line) after being normalized by the continuum. The measured limits of the BAL are shown as vertical red lines. The horizontal blue dotted line represents a normalized flux of 1.0, and the horizontal red line shows a normalized flux of 0.9. } 
\label{fig:het_bal} 
\end{center} 
\end{figure}

We here provide some additional details on measurements made from the HET spectrum, which was obtained during poor weather conditions and analyzed separately from the set of SDSS spectra. We measure SNR$_{1700}~=~4.74$ in this spectrum, which is lower than most of our SDSS spectra but still high enough for us to obtain useful constraints. We processed this spectrum in a similar fashion as the SDSS spectra; we first cropped and cleaned/interpolated the data, converted it to the quasar rest frame, and fit a reddened power law continuum. Because of the lower signal, we focused only on the high-velocity Trough A BAL and did not make measurements of Troughs B and C; we thus did not fit the \civ emission line in this particular case. Figure~\ref{fig:het_bal} shows the continuum-normalized HET spectrum, focused on the \civ region.  

Interestingly, Trough A appears to have weakened substantially between our last SDSS observations (in 2021) and the new spectrum --- in addition, it has returned to having two distinct sub-troughs. In fact, it is now weak enough that the two sub-troughs appear as two separate troughs, with the flux between the two of them recovering to the continuum level. For consistency with our previous measurements, we measure the BAL properties across the entire feature, from the blueward limit of the blue sub-trough to the redward limit of the red sub-trough --- from 1444.95\,\AA\ and 1470.75\,\AA. Within this region, we measure an EW of 3.34$\pm$0.27\,\AA, a mean depth $d$~=~0.13$\pm$0.01, and $v_{\rm cent}$~=~$-18,110$~km~s$^{-1}$ (Table~\ref{Table:mean_BAL_measurements}). These measurements show that the BAL has weakened dramatically since the SDSS monitoring, and has continued to shift in the blueward direction (the blue and red BAL boundaries were roughly around 1446\,\AA\ and 1472\,\AA\ at the very end of Year 8; the BAL has thus continued to drift blueward by roughly 1--2 \AA\ in the last two years). 

\section{Model fits to Trough A Sub-Troughs} 
\label{app:modelfits}

\begin{figure}
\begin{center}
\includegraphics[scale = 0.28, trim = 0 0 0 0, clip]{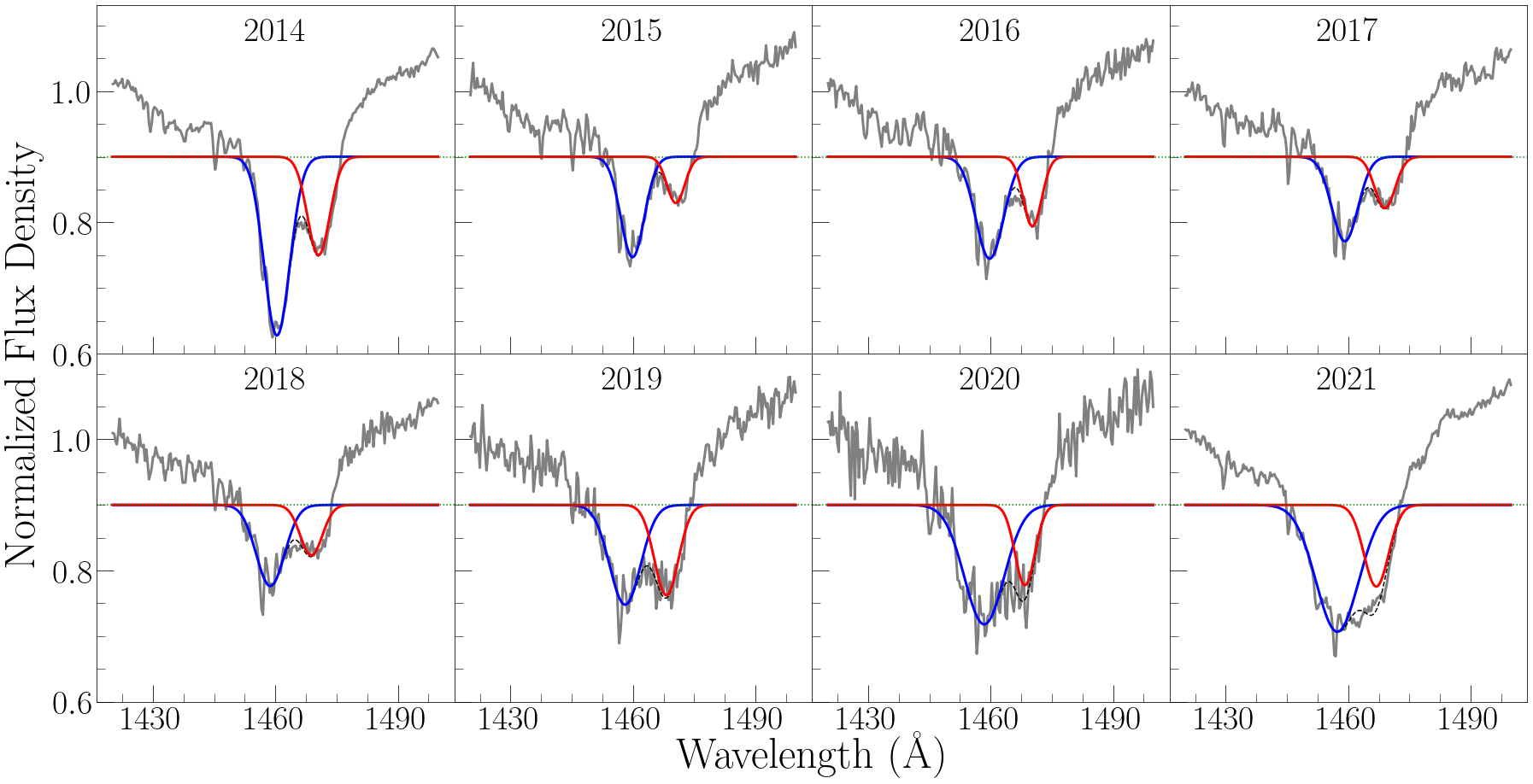} %gaussian_fits.png
\caption{The two-Gaussian model fits to the two sub-troughs within Trough A for the mean spectra of each individual year of monitoring. Each mean spectrum is shown in gray; red shows the redward Gaussian component of the model fit to the data and blue represents the blueward Gaussian component. } 
\label{fig:subtrough_fits} 
\end{center} 
\end{figure}

Section~\ref{sec:subtroughs} presents a discussion of model fits to the two distinct sub-troughs of Trough A. We adopt a simple model, using a single Gaussian component for each sub-trough. Figure~\ref{fig:subtrough_fits} shows these model fits, demonstrating the evolution of the two distinct sub-troughs over time. 

\bibliographystyle{aasjournal} 
%\bibliography{paper_refs}

%/include{bibliography}

\end{document}